\documentclass[12pt]{article}

\pdfoutput=1
\usepackage{jheppub}
\usepackage{amsmath,amssymb,amsthm}
\usepackage{graphicx,subfigure,slashed}

\newcommand{\beq}{\begin{eqnarray}}
\newcommand{\eeq}{\end{eqnarray}}

\newcommand{\met}{{\slashed E_T}}

\begin{document}

\title{Unbroken $SU(2)$ at a 100~TeV collider}

\author[a]{Anson Hook,}

\author[b]{and Andrey Katz}

\affiliation[a]{School of Natural Sciences, Institute for Advanced Study, Princeton, NJ 08540}

\affiliation[b]{Center for the Fundamental Laws of Nature, Jefferson Physical Laboratory,\\ 
Harvard University, Cambridge, MA 02138}

\abstract{A future 100~TeV pp collider will explore energies much higher than the scale of electroweak (EW) symmetry 
breaking. In this paper we study some of the phenomenological consequences of this fact, concentrating on 
enhanced bremsstrahlung of EW gauge bosons. We survey a handful of possible new physics experimental searches one 
can pursue at a 100~TeV collider using this phenomenon.
The most dramatic effect is the non-negligible radiation of EW gauge bosons from neutrinos, 
making them partly visible objects.  The presence of collinear EW radiation allows for the 
full reconstruction of neutrinos under certain circumstances.  We also show that 
the presence of EW radiation allows one 
to distinguish the $SU(2)$ quantum numbers of various new physics particles. 
We consider examples of two completely different new physics paradigms, additional gauge groups and SUSY, where the bremsstrahlung
radiation of $W$ and $Z$ from $W'$s, $Z'$s or stops allows one to determine the couplings and the mixing angles 
of the new particles (respectively).  
Finally, we show how the emission of $W$s and $Z$s from high $p_T$ Higgs bosons can be 
used to test the couplings of new physics to the Higgs boson.}

\maketitle

\section{Introduction}

Every new collider has given us valuable information about the structure of the Universe. The recent discovery
of the Higgs boson by the ATLAS~\cite{ATLAS:2012ae} and CMS~\cite{Chatrchyan:2012tx} collaborations 
is the most recent example of this trend.  On the other hand,
the LHC still has not discovered any sign of the much anticipated new physics. These negative results are perplexing as naturalness 
considerations clearly favor new physics at the electroweak (EW) scale. Of course, it is possible that the current 
searches have simply overlooked new physics hidden inside of the enormous QCD backgrounds and new exciting
discoveries are right around the corner. It is also possible that spectacular signals will show up in the early LHC14 data. 
The current bounds on colored new physics push the generic new physics scale to the TeV scale, already 
in tension with naturalness. 
If nothing is found below a few TeV,
this would indicate that the Universe is not perfectly natural and is probably guided by principles other 
than naturalness. 

If we find no new physics in the upcoming LHC14 run, 
it would very important to know how unnatural the world is. If it is only tuned to the percent level, then it is a tuning we 
have seen before in nature  and naturalness still works as a guiding principle.  If it is tuned beyond the percent level, 
we would need to find new guiding principles.
  
On the other hand, if new physics is found at the LHC14 or even in earlier data, it would be important 
to perform precision measurements of the new physics.  Both of these goals are very 
well served by a future 
high energy hadron collider with $\sqrt{s} = 100$~TeV.  This future collider would be able to 
extend the reach for new colored particles 
to the range of dozens of TeV, potentially discovering the new physics responsible for  the 
``almost naturalness'' of the EW scale physics.
On the other hand, if new physics is found at the LHC14, then the 100~TeV future machine 
would be an excellent tool with which one can  perform precision measurements on these new particles.  

In this paper we explore some of the new and surprising aspects of a 100~TeV collider. 
Most importantly, we demonstrate that a 100~TeV machine is not a simple rescaling of lower-scale pp colliders.  
To this end, we focus on a new effect that just starts to become important at a 100 TeV~collider, namely that if built, 
it will be the first machine ever where the typical energy of interactions \emph{is much higher than the EW symmetry 
breaking scale}. Therefore, in these collisions the EW symmetry $SU(2)_L \times U(1)_Y$ can be effectively treated as 
unbroken. This simple observation manifests itself in several non-trivial phenomena that are largely
 inaccessible 
at the LHC or at any other lower energy collider.   

The most important effect  of the EW force being a ``long-range force'' at the 100~TeV collider is the 
enhancement of the EW radiation of $W$ and $Z$ bosons. The Sudakov double-log 
enhancement of photon and gluon emission at lower energies has been well studied.  Consider a particle of mass $m$
charged under the electromagnetic force emitting a photon. 
Let the incoming momentum of the particle be $p$  and the outgoing momentum be $p'$. 
The differential 
cross sections for this process reads
\beq\label{Eq:sudakov}
d \sigma (p \to p' + \gamma) \approx d\sigma (p \to p') \times \frac{\alpha}{\pi} 
\log \left( - \frac{(p - p')^2}{\mu^2} \right) \log \left( - \frac{(p-p')^2}{m^2}\right)~.
\eeq  
This is the well known result of Sudakov double log enhancement. The first logarithm in this expression 
is an IR divergence and it is cut off by the IR cut off $\mu$ (which for the photon comes about from our inability to 
detect arbitrarily soft photons and for electroweak radiation is the mass of the $W$ and $Z$ bosons).  
The second log is a collinear singularity, which is cut off by the 
mass of the emitting particle $m$.\footnote{Of course, in practice to take into account correctly the emission of photons (gluons)
in QED (QCD) one cannot just rely on~\eqref{Eq:sudakov}, 
since the ``subleading" terms are not longer small compared to the leading
one. As we will see, the effect of EW radiation in 100~TeV, although appreciable, is not that strong, and it does not 
demand a full resummation. }  

At multi-TeV energies, the analogous process happens 
with any particle charged under $SU(2)_L$. A quark with energy $E \gg v$ will have a probability to emit
$W$ and $Z$ bosons in agreement with the above mentioned formula, up to corrections due to the non-Abelian 
nature of the 
EW force (for relevant works involving EW Sudakovs, 
see Refs.~\cite{Ciafaloni:2006qu,Baur:2006sn,Bell:2010gi,Campbell:2013qaa}).  
In this case, the IR cutoff naturally becomes $\mu = m_W, m_Z$. Note that regardless 
of the mass of the emitting particle, $W/Z$ emission at high energies is always enhanced by a single log 
due to the IR singularity.  If the emitting particle is light, then the enhancement is double-log due to the collinear
singularity.  However, if the new heavy particles are at the TeV scale, then the colinear singularity 
is cut off by the mass of the heavy particle and 
EW radiation is only single-log 
enhanced.

In this paper we consider three novel and exciting applications for EW radiation in the search for new physics.  
The first is the idea of neutrino tagging, which is most clearly illustrated in the example of a new heavy $Z'$ or $W'$ 
boson.\footnote{The importance of heavy $Z'$ three-body decays was 
first mentioned in Ref.~\cite{Cvetic:1991gk}
in the context of SSC and later in Ref.~\cite{Rizzo:2014xma} in context of a 100~TeV collider,} 
Neutrinos are charged under $SU(2)_L$ and will therefore 
emit $W$ and $Z$ bosons when produced at large $p_T$s.  
The production of the EW gauge bosons is both IR and colinear-enhanced.   
The collinear singularity results in the reconstructed $Z$ boson being 
strongly preferred to be collinear with the neutrino, up to small corrections due to the ``dead zone''.  
Assuming that a neutrino lies almost parallel 
to a reconstructed $Z$ allows one to reconstruct missing energy in events where there is only one 
neutrino giving the missing energy, e.g. a $W'$ event.  A more dramatic example is if a $W$ boson is radiated.  
The neutrino 
becomes completely visible and instead it becomes important to tag the origin of the lepton as a neutrino.

Another use for EW radiation is to identify the quantum numbers of new particles, both visible and 
invisible. Most of the Standard Model (SM) production, even at 100~TeV collider is near threshold, 
such that the effect of the EW radiation is often minor.  
However, when new particles are produced in a cascade, they are typically produced at high $p_T$ and 
thus have a high probability of radiating a $W$ or a $Z$ boson.
We take advantage of this fact and consider two different SUSY spectra, which are examples of new particles produced 
in cascade decays. We show that both these spectra enable extraction of the quantum numbers of various particles 
based on the EW radiation pattern.   
As cascade events are typically messy, 
we focus on the enhancement of the total cross section rather than any soft or collinear singularity.  
We show that the total cross section varies by an order of magnitude as the quantum numbers of particles 
are varied.  Thus the measurement of additional $W$ and $Z$ radiation can provide supplemental information 
regarding the quantum numbers of new particles.

Finally, we demonstrate that the EW radiation coming from high-$p_T$ $W$, $Z$, 
and Higgs bosons can be used to constrain the couplings of new physics to the Higgs boson. 
We consider  a new scalar which couples to the SM-like Higgs.
Much like how $WW$ scattering can be used to probe the Higgs couplings to the $W$ boson, 
the branching ratios of this new particles
shed light on the couplings in the Higgs sector. 
We point out that the three body branching ratios of the new scalar
are even more sensitive than the two body branching ratios to the value of these couplings.
Much like $WW$ scattering, the three body branching ratios can grow faster than logarithmically
if the $WW$  scattering is not \emph{fully} unitarized just by the SM-like Higgs boson.  
Thus they provide a unique probe of the coupling of new physics to the Higgs boson.

Throughout this paper, we consider physics at the partonic level without any detector simulation.  As we do not yet know what the detector specifics of a 100~TeV machine would be, we cannot reliably model detector effect.  
As such, we expect our results to be correct only up to $\mathcal{O}(1)$ numbers.  However, as all of our 
effects have a sound physical basis, we expect that a more detailed account of the physics would not over 
turn any of the results.  Our motivation is simply to demonstrate that there are new and exciting 
effects at 100~TeV (for other interesting new 100~TeV ideas see Ref.~\cite{Larkoski:2014bia}).

Our paper is organized as follows. 
In Sec.~\ref{Sec:ZWprime}, we study the radiation of $W$ and $Z$ bosons from a heavy $W'$ or $Z'$ 
gauge boson.  
In Sec.~\ref{Sec:quantum}, we investigate the use of electroweak radiation in
precision measurements of TeV-scale SUSY particles. In particular, we show that EW gauge boson emission 
can give us information about the quantum numbers of the LSP under $SU(2)_L \times U(1)$ as well as the mixing
angles of an NLSP stop. 
In Sec.~\ref{Sec:unitarity}, we demonstrate how three body branching ratios of a new physics particle provide 
a unique probe of the couplings in the Higgs sector.  
In Sec.~\ref{Sec:conclusion} we conclude and comment on more 
possible searches one can perform at 100~TeV machine along these lines. 
Finally, in the appendix we briefly comment on the potential of the 100~TeV collider to determine the quantum numbers of 
SUSY DM without cascade decays.


\section{Seeing the invisible - a $W'/Z'$ case study}
\label{Sec:ZWprime}


The invisible and semi-invisible decays of a $Z'$ and $W'$ are difficult to probe directly. 
On the other hand, since 
any $Z'$ (unless extra exotic matter is introduced) is expected to be a 
linear combination of hypercharge and $B-L$ so we expect that it should have an appreciable 
invisible two-body decay rate $Z' \to \nu \bar \nu$.  The exact rate will depend on the mixing angle
between $U(1)_Y$ and $U(1)_{B-L}$, something that we will loosely call the chirality of the $Z'$.  
On the other hand, a BSM $W'$ can have a semi-invisible two-body 
decay mode $W' \to l \nu$. 
  
At large energies, neutrinos can emit $W$ and $Z$ bosons making missing energy visible. 
The double-log  enhancement of this process can make the three-body decays of a $W'$ or $Z'$ significant
if the leptons are sufficiently boosted,
e.g. $Z' \to \nu \bar \nu Z$ or $Z ' \to \nu l^- W^+$.  
These three-body decays contain important information on the  
couplings of the $Z'$ and the mixing angle between $U(1)_{B-L}$ and $U(1)_Y$.\footnote{This issue was 
first explored in the context of the SSC in Ref.~\cite{Cvetic:1991gk}}  
The $p_T$ of the primary leptons or neutrinos are proportional to the mass of the heavy bosons and for 
$Z'$s that are kinematically accessible at the LHC, these rates are too small to be observed. On the other hand, at a future 
100~TeV collider these decay modes would be very important.

The radiation of a $W$ or $Z$ from a neutrino has a soft 
collinear log enhancement, which is cut off by $m_{W, Z}$.  If a $Z$ boson is radiated, 
the collinear enhancement results 
in a strong tendency 
for the $Z$ boson to be emitted parallel to the neutrino.  
Assuming that the $Z$ boson lies completely 
parallel to the neutrino allows one to 
reconstruct the neutrino in its entirety.  If a $W$ boson is radiated, 
the neutrino becomes completely visible and tagging its origin as a 
neutrino is needed.  If the $W$ can be reconstructed (most likely in a hadronic decay mode), 
the small $\Delta R$ distance between it and the lepton allows one to tag the lepton as originating from a neutrino.  

The soft divergence results in gauge bosons which are often not significantly boosted.  
Thus their decay products can be widely separated.
Although tagging of hadronic $W$s and $Z$s is possible at the LHC (although this usually comes at a price of an appreciable
mistag rate), 
it is not clear how feasible it will be  
at future hadronic colliders, especially as we do not know the details of the detectors and QCD backgrounds (for a review of 
boosted gauge boson hadronic tagging at the LHC see e.g.~\cite{Altheimer:2012mn}).
Therefore
in this work we will concentrate on the low BR, but spectacular and low-background all 
leptonic decays of the $Z$.  
The advantage is that the entire $Z$ can be easily reconstructed.  
We signify a leptonically decaying $Z$ 
by the notation $Z_l$.
  
We illustrate the utility of leptonically decaying EW-radiated $Z$ bosons in two examples. 
The first example is a heavy $W'$ boson.  When looking at events with an extra $Z$, 
the tendency of the $Z$ to lie parallel with the neutrino allows for the 
reconstruction of missing energy rather than missing transverse energy.  The reconstructed neutrino 
can then be used to find the mass of the decaying $W'$ boson. We show that in this case a full reconstruction
of the $W'$ mass peak is possible.

The second example is a $Z'$ boson.  The magnitude of the invisible channel can be probed by the emission of a $W$ or $Z$ boson.  
Probing the invisible channel allows one to directly measure the $Z'$s couplings to the left handed leptons.

\subsection{$W' \to l \nu Z_l$}

In the decay of a $W'$ particle, the neutrino can be tagged by the collinear radiation of a $Z$.  Tagging the direction of a neutrino by assuming that the neutrino is collinear with the radiated $Z$ allows 
one to reconstruct the neutrino four vector momentum in its entirety.  This in turn allows us to reconstruct the 
missing energy, not just the missing transverse energy, and therefore reconstruct the mass of the $W'$.  
Of course the $Z$ and the neutrino are not perfectly parallel to one another, 
so even before detector effects are taken into account,
we end up with a smeared resolution for the $W'$ mass.  

 \begin{figure}
\centering
\includegraphics[width=0.47\linewidth]{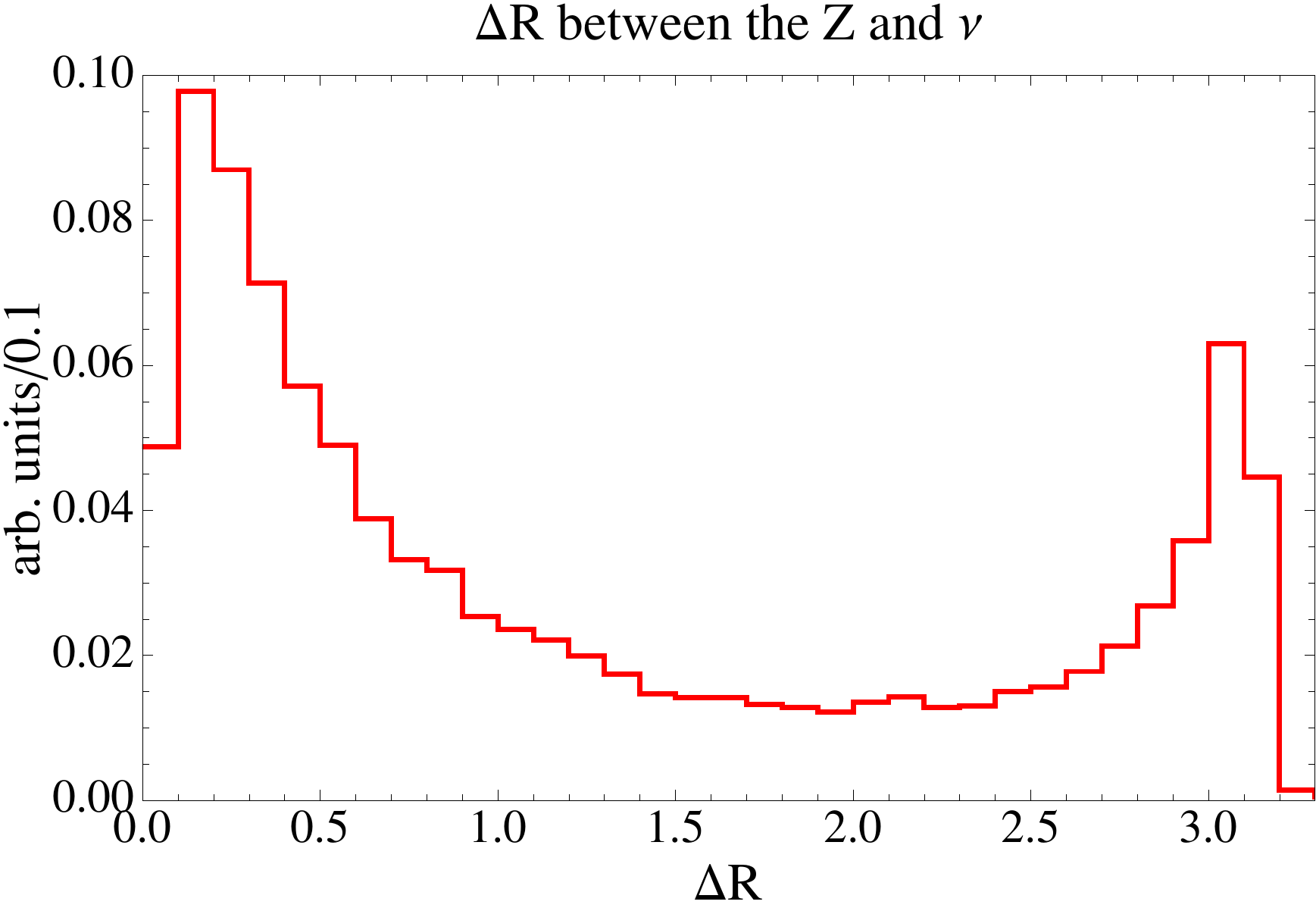}
\includegraphics[width=.47\linewidth]{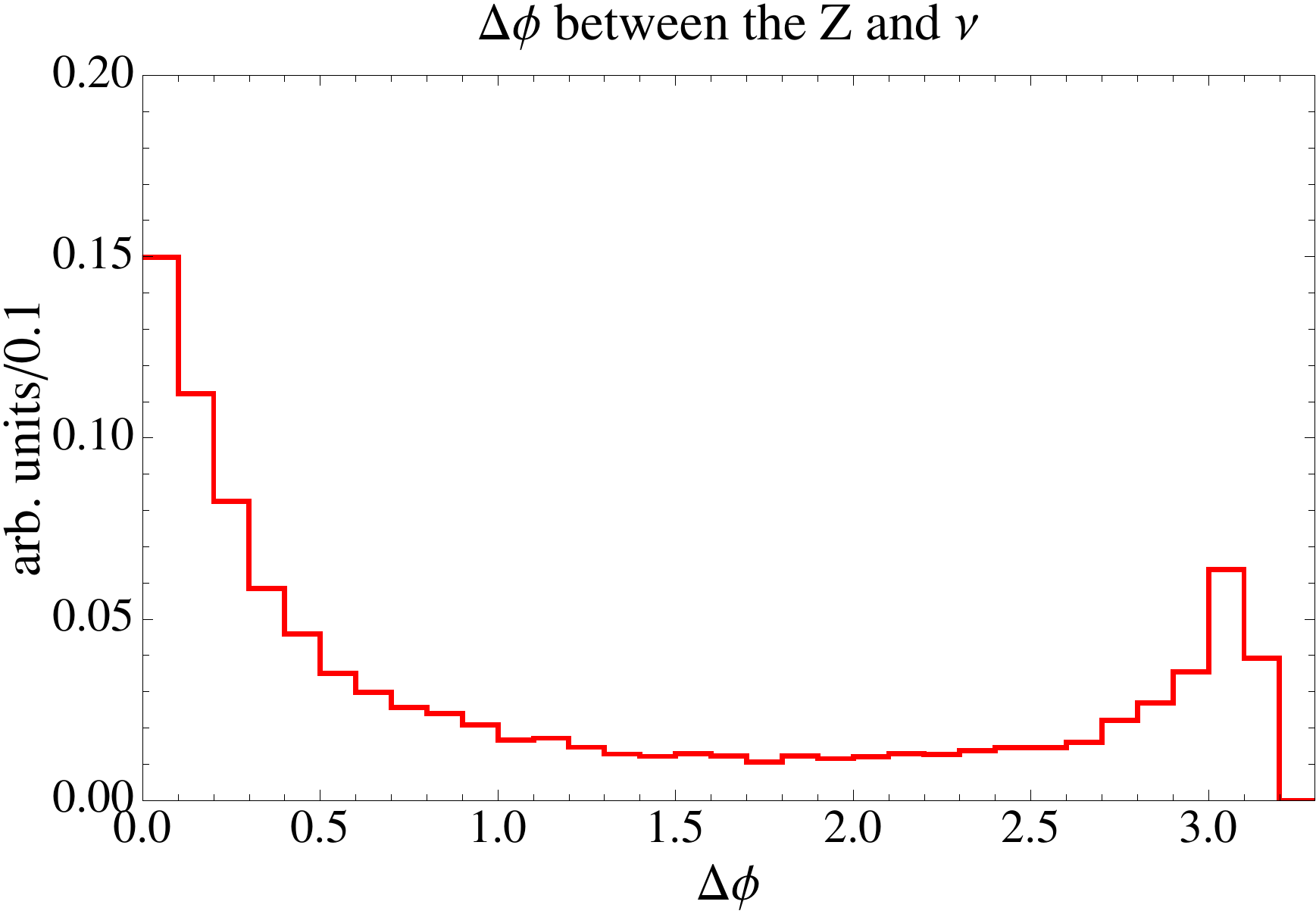}    
\caption{ $\Delta R$ and $\Delta \phi$ between the neutrino and the $Z$ for a 5~TeV $W'$ 
decaying to a lepton, neutrino and a $Z$.  EW radiation has a collinear singularity cut off by the 
mass of the $W/Z$ bosons.  The effect of this collinear enhancement is clearly seen.  The $Z$ is radiated 
off of either the neutrino or the lepton.  Given that the neutrino and lepton tend to be back to back 
there is an enhancement 
at $\Delta R$ or $\Delta \phi \approx 3$ due to the $Z$s radiated off of the visible lepton.
Since the $Z$ couples to the neutrinos stronger than to the leptons, this enhancement is smaller 
than at $\Delta R \approx \Delta \phi \approx 0$. }  
\label{Fig: dR}
 \end{figure}

To illustrate these effects, we first consider a 5~TeV $W'$ boson. Along with the dominant $l \bar \nu$ decay mode, 
it is expected to have a subdominant but appreciable 3-body decay mode $l \bar \nu Z$ that we will be interested in.  
To eliminate any standard model backgrounds, we first place a 500~GeV $p_T$ cut on 
the leading lepton, the missing 
transverse energy and the transverse mass of the $\met$ and the leading lepton.\footnote{This cut is sufficient 
to eliminate most of the 
dominant background, namely $WZ+$jets. While we do not estimate what portion of this reducible 
background will survive after full hadronization and detector simulation.  We do not expect it to be significant and further
improvement will become feasible when we discover more about the detectors of a 100~TeV collider.  
Another background is $W_l Z_lZ_{inv}$ but its cross section is extremely small.}  
A 5~TeV $W'$ has very high energy leptons and neutrinos leading to large double-log Sudakov factors and therefore 
non-negligible three-body decay rates.   

The additional $Z$ in these events can come from ISR, radiation off of the $W'$ and FSR from both the lepton and neutrino. 
The last is of course especially
interesting for us as we are interested in genuine three-body decays where the $Z$
is expected to be roughly collinear with the neutrino or lepton.
To show the effect of the collinear log enhancement, we plot in Fig.~\ref{Fig: dR} 
the  distribution of  $\Delta R$ and $\Delta \phi$ between the reconstructed $Z$ and the neutrino.\footnote{As usual, we define $\Delta R \equiv \sqrt{\Delta \eta^2 + \Delta \phi^2}$.}  
The collinear enhancement is seen very clearly.  The $Z$ has larger couplings to the neutrino than to the leptons as can also be seen in the plot as the lepton and neutrino are roughly back to back.

When the $\Delta R$ between the neutrino and the $Z$ is small, then the direction of the $Z$ 
approximately corresponds to the spatial direction of the neutrino, 
thus allowing the full reconstruction of the latter.   
To establish that the leptonic $Z$ was Sudakov radiated off of the neutrino rather than the 
lepton, we put a $\Delta \phi_{Z_l \met} < 0.5$ 
cut between the 
reconstructed $Z$ and the missing energy.  $Z$s emitted from ISR which happen to point in the same $\phi$ direction as the missing energy 
can be effectively removed by requiring that \emph{the reconstructed Z boson} has $|\eta| < 2.5$ 
(not to be confused with the 
acceptance cut that we put on the leptons 
themselves).

 \begin{figure}
\centering
\includegraphics[width=0.44\linewidth]{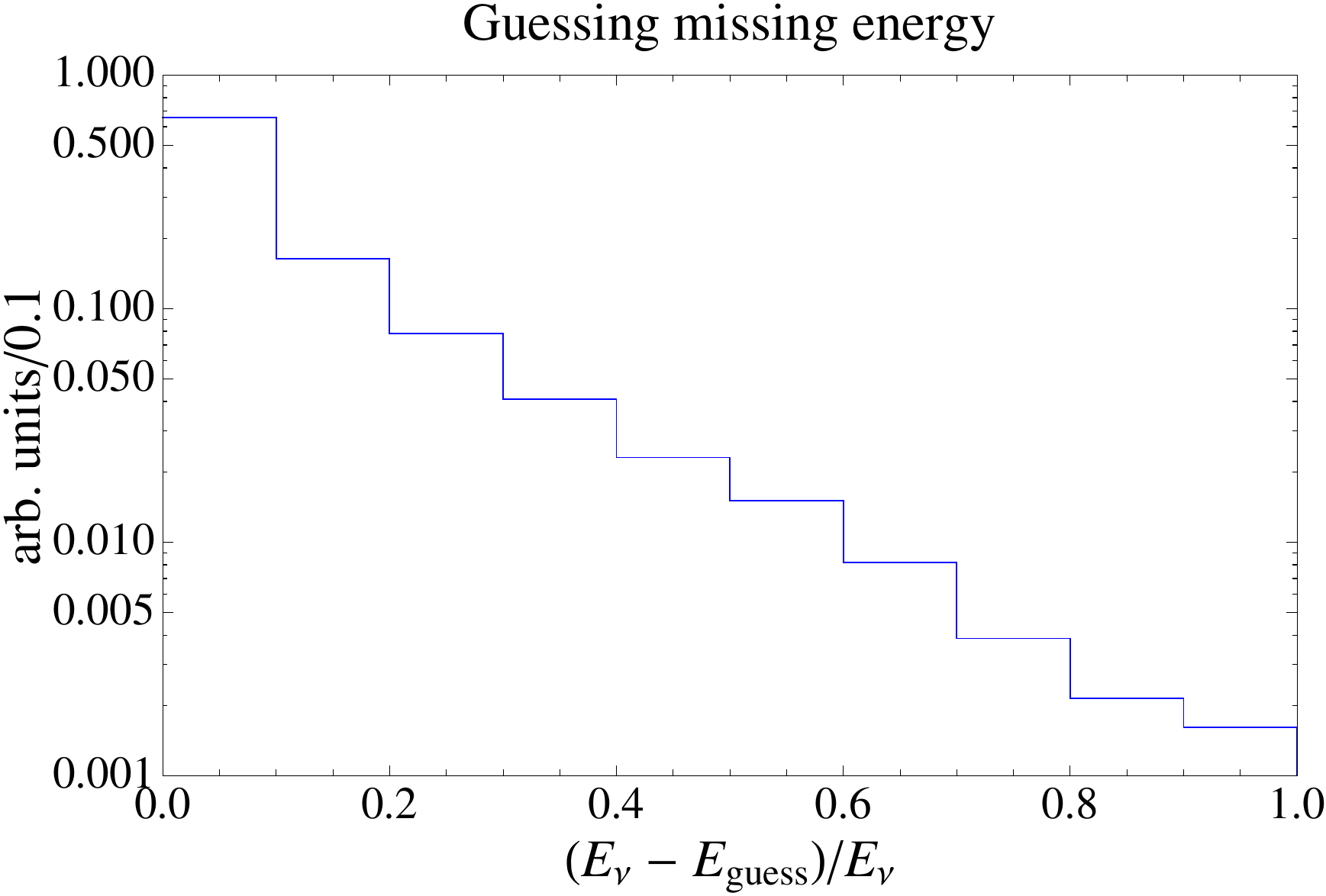}
\includegraphics[width=0.44\linewidth]{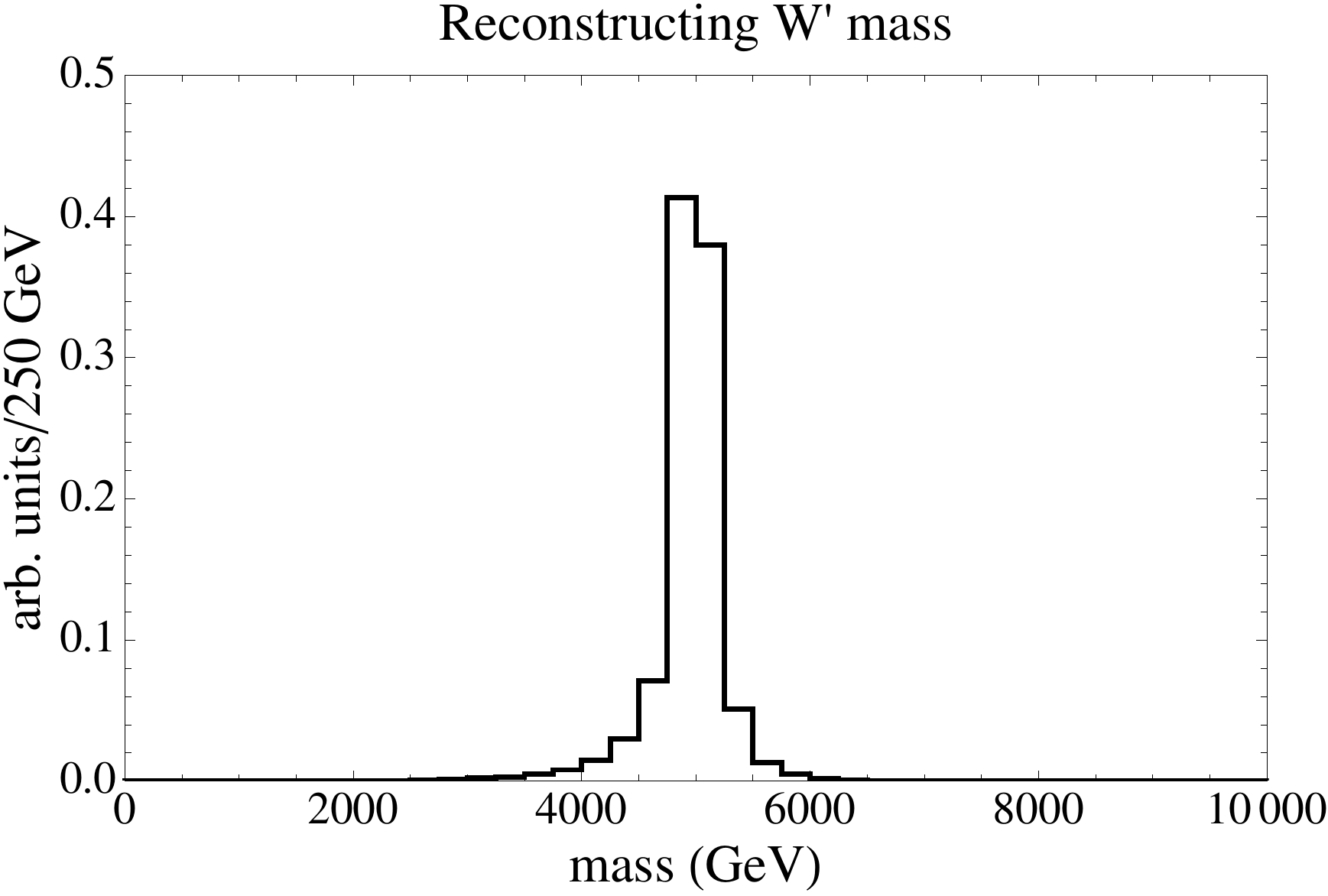}
\caption{ On the LHS we plot the difference between our ``guess'' about the energy 
of the neutrino and the actual neutrino energy.  The ``guess'' for the neutrino energy comes 
from assuming that the neutrino is perfectly collinear with the leptonic $Z$. 
The reconstructed $Z$ is required to have $|\eta| < 2.5$ and $\Delta\phi_{Z \slashed{E}_T} < 0.5$.  
The reconstructed neutrino allows one to guess the real missing energy in an event as well as reconstruct the 
full mass peak of a $W'$ particle (plot on the RHS). The mass resolution is smeared since the Z is not 
always collinear with the neutrino, but there is a very clear peak at the $W'$ mass of 5~TeV.}  \label{Fig: wprime}
 \end{figure}

We work at parton level assuming that the leptons and missing energy are measured perfectly.  Madgraph5~\cite{Alwall:2011uj,Stelzer:1994ta,Maltoni:2002qb} was used to generate the events.   
In this very preliminary analysis, alongside with the standard acceptance criteria, we apply following cuts:
\begin{itemize}
\item Exactly three leptons (either $e$ or $\mu$) in the event
\item $p_T > 0.5$~TeV for the leading lepton
\item The invariant mass of the subleading leptons reconstructs the Z mass.  $75$~GeV~$<m_{ll} < 105$~GeV
\item Eta of all leptons and reconstructed Z obey $|\eta_{Z,l}| < 2.5$
\item $\Delta \phi_{Z \met} < 0.5$
\item $\met > 0.5$~TeV and $m_T(l,\met) > 0.5$~TeV
\end{itemize}
 
We present the results of the $W'$ reconstruction in Fig.~\ref{Fig: wprime}.  
In most of the cases, the missing 
energy can be reconstructed to the 
precision of $\sim 20\%$
or even better.  The reconstructed mass peak (the same figure on the right) is a little smeared due to imperfect reconstruction of the neutrino, 
however the mass peak is still clearly visible.  For a 5~TeV $W'$, $\sigma \times Br \times \epsilon \sim 14 $ fb.  Thus for a rather reasonable integrated luminosity, we can obtain enough signal events to easily determine the mass of the $W'$ through this method.  The dominant background comes from $WZZ$ 
which has a negligible cross section times efficiency.

\subsection{$Z' \to \nu \bar \nu Z_l $}
\label{Sec:Zprime}

Discovery of a $Z'$ at a hadron collider is simple unless it is completely leptophobic.
On the other hand, certain decay modes are considered to be hard or almost impossible to measure. A canonical example 
of such a decay mode is a completely invisible decay $Z'\ \to \nu \bar \nu$. Of course this stays true at a 100~TeV machine.  
However if the $Z'$ is heavy enough, the existence of the invisible mode inevitably implies the existence of a subdominant decay mode 
$Z' \to \nu \bar \nu Z$. The ratio between two these modes
\beq\label{eq:Zfromnu}
\frac{\Gamma(Z' \to \nu \bar \nu Z)}{\Gamma(Z' \to \nu \bar \nu)}
\eeq 
is determined by the rate of 
the EW radiation and therefore \emph{only depends on the mass of the $Z'$ and not on the couplings 
of the $Z'$ to the chiral fermions}. If this 
3-body decay mode is abundant enough and can be clearly detected, we can determine the 
number of events in the invisible channel.

Even more important, if the $Z'$ is discovered, measurement of the invisible mode allows us to determine 
the couplings of the $Z'$ and the mixing angle between the $U(1)_Y$ and $U(1)_{B-L}$  
When EW radiation is present, the coupling to the chiral leptons is easy to determine.  
Neutrinos are purely left handed particles so measuring the invisible channel gives a direct probe of the 
$Z'$s coupling to the left handed leptons.
  
As in the previous case, we take advantage of a subdominant but spectacular leptonic decay 
mode of the $Z$.  It would be interesting to study in future whether one can exploit the hadronic 
decay modes of the $Z$. The characteristic signature of this decay is $(Z \to l^+ l^- )+ \met$. 
The main background that we consider for this process is $(Z\to l^+ l^-) (Z\to \nu \bar \nu)$. Other subdominant 
backgrounds include $W Z$ (when the lepton from the $W$ decay is lost), $t \bar t$ and $W^+ W^-$. 
The last two backgrounds can be efficiently removed by 
an appropriate cut on $m_{T2}$. $t \bar t$ near threshold can be even further reduced by imposing a jet veto. Therefore, we 
neglect these subdominant backgrounds and estimate our reach by comparing the signal to $Z_{ll}Z_{inv}$.
 
Parenthetically  it is interesting to note that at high energies $ZZ+$jets is expected to
 be \emph{significantly smaller}
than other EW backgrounds like $WW+$jets and $WZ+$jets. The reason for this is very simple: at a 100~TeV 
machine the later 
backgrounds are dominantly produced when one EW gauge boson radiates off 
of another one with the entire system recoiling against the 
hard jet(s).  This particular kinematic configuration is Sudakov (double-)log enhanced.  
However, a $Z$ cannot be radiated from another $Z$, which leaves the $ZZ$+jets backgrounds without this important 
enhancement and thus subdominant to the other backgrounds.

\begin{figure}
  \centering
  \includegraphics[width=0.44\linewidth]{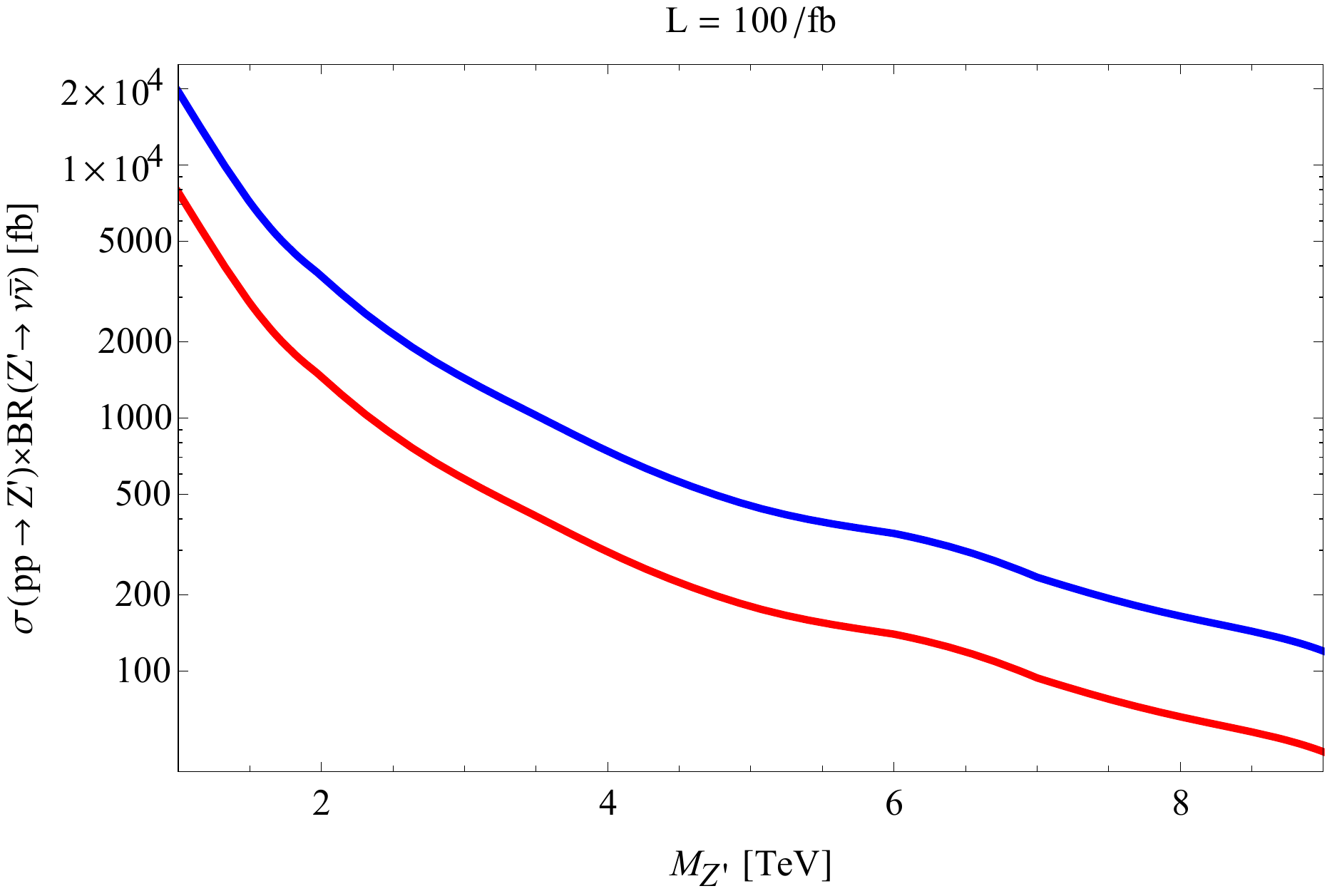}
  \includegraphics[width=0.44\linewidth]{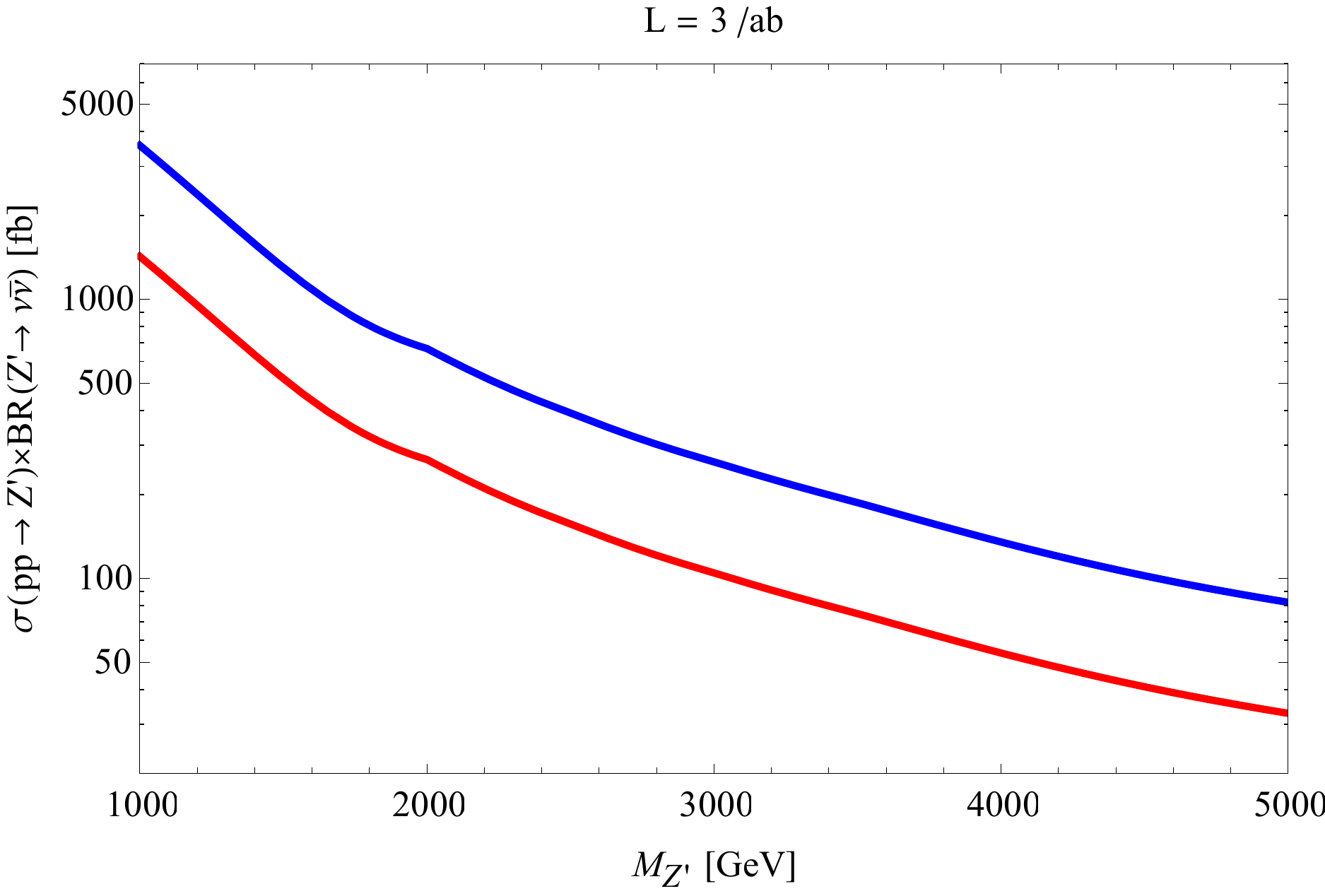}
 \caption{Reach of a 100 TeV collider to a $Z'$ decaying invisibly for a luminosity of 100 fb$^{-1}$ and 3000 fb$^{-1}$
as extracted from measurement of $Z' \to \nu \bar \nu Z$ channel.  
The blue and red lines are the 5 and 2 $\sigma$ results respectively.}  
\label{Fig: Zprime}
 \end{figure}

Unlike in the $W'$ case, full reconstruction is of course not feasible. 
Therefore we perform a simple cut-and-count search for leptonic $Z$ 
recoiling against $\met$.  We compare the rate of these events after acceptance cuts and a cut 
on $\met$ to the rate of the $ZZ$ background to determine the possible reach of the 100~TeV machine.  
The results of this search are shown in Fig.~\ref{Fig: Zprime}.  We phrase our results in terms of 
\beq
\sigma(pp \to Z' ) \times BR(Z' \to \nu \bar \nu)
\eeq
as the ratio~\eqref{eq:Zfromnu} is known for every  given mass of the $Z'$.   
As we see from the figure, the future hadron collider can probe
invisible decays of $Z'$s with couplings of order ${\cal O}(0.1)$.

Let us now turn to the interpretation of these results. 
There are two anomaly free symmetries in the Standard Model,\footnote{We assume universal 
couplings to the three generations.} 
$U(1)_{B-L}$ and $U(1)_Y$.  In the minimal scenario, with no exotic matter fields, 
a $Z'$ can couple to these two $U(1)$s as 
\beq
g (\sin\theta \ U(1)_Y + \cos\theta \ U(1)_{B-L})~. 
\eeq  
The mixing angle fixes the couplings of the $Z'$ to the chiral fermions of the SM.  

\begin{figure}
\centering
\includegraphics[width=0.44\linewidth]{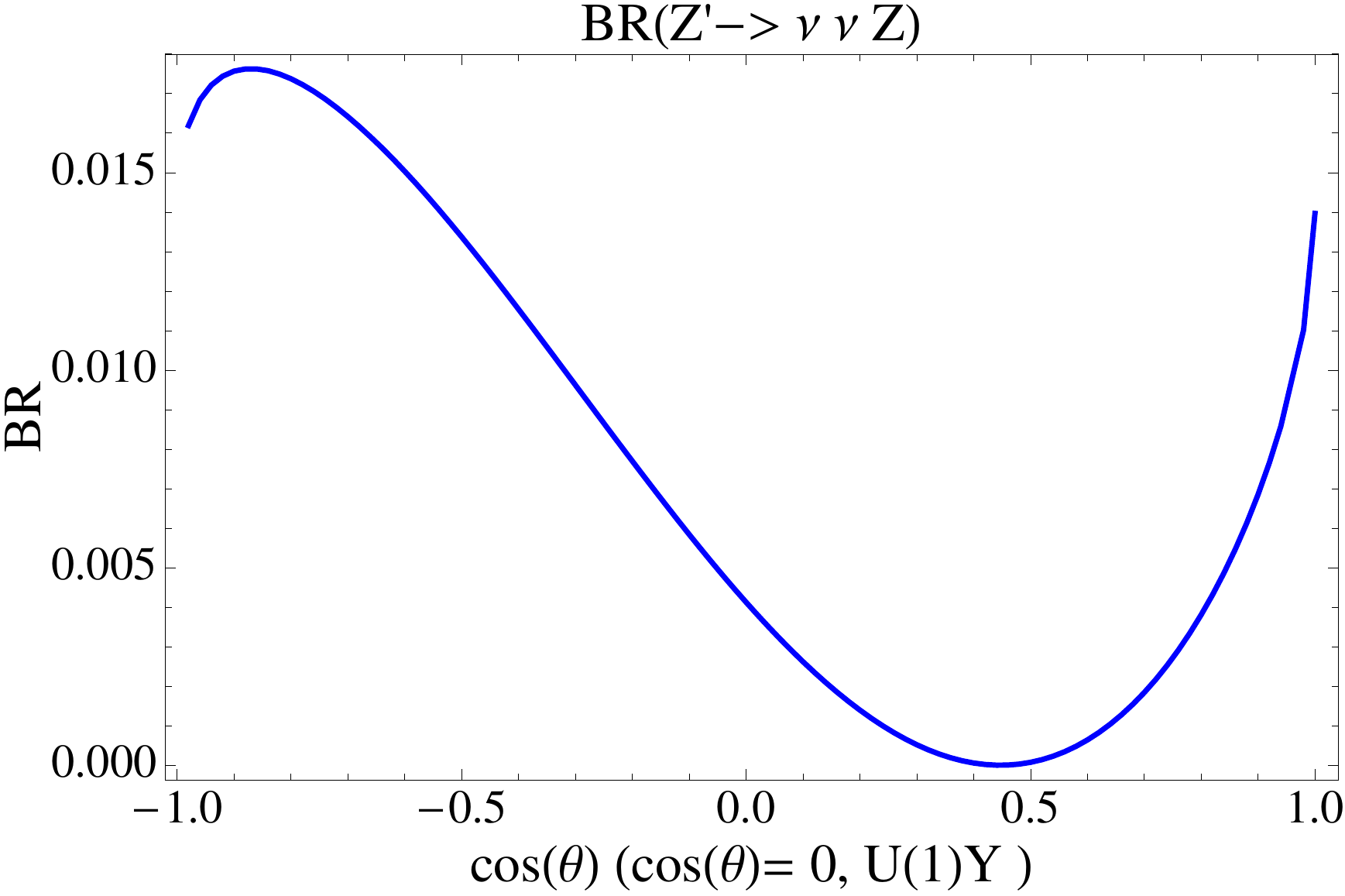}
\includegraphics[width=0.44\linewidth]{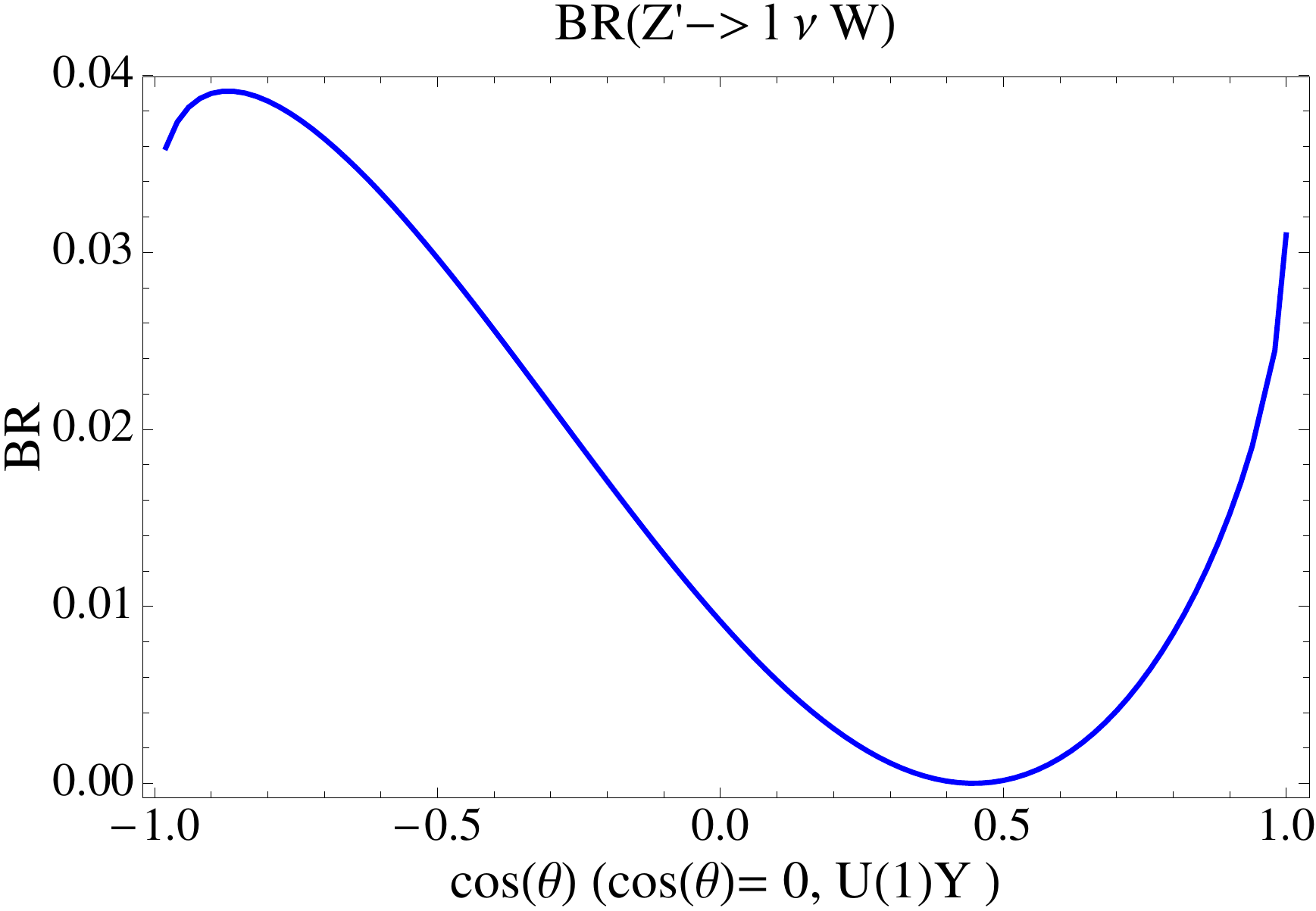}
\includegraphics[width=0.44\linewidth]{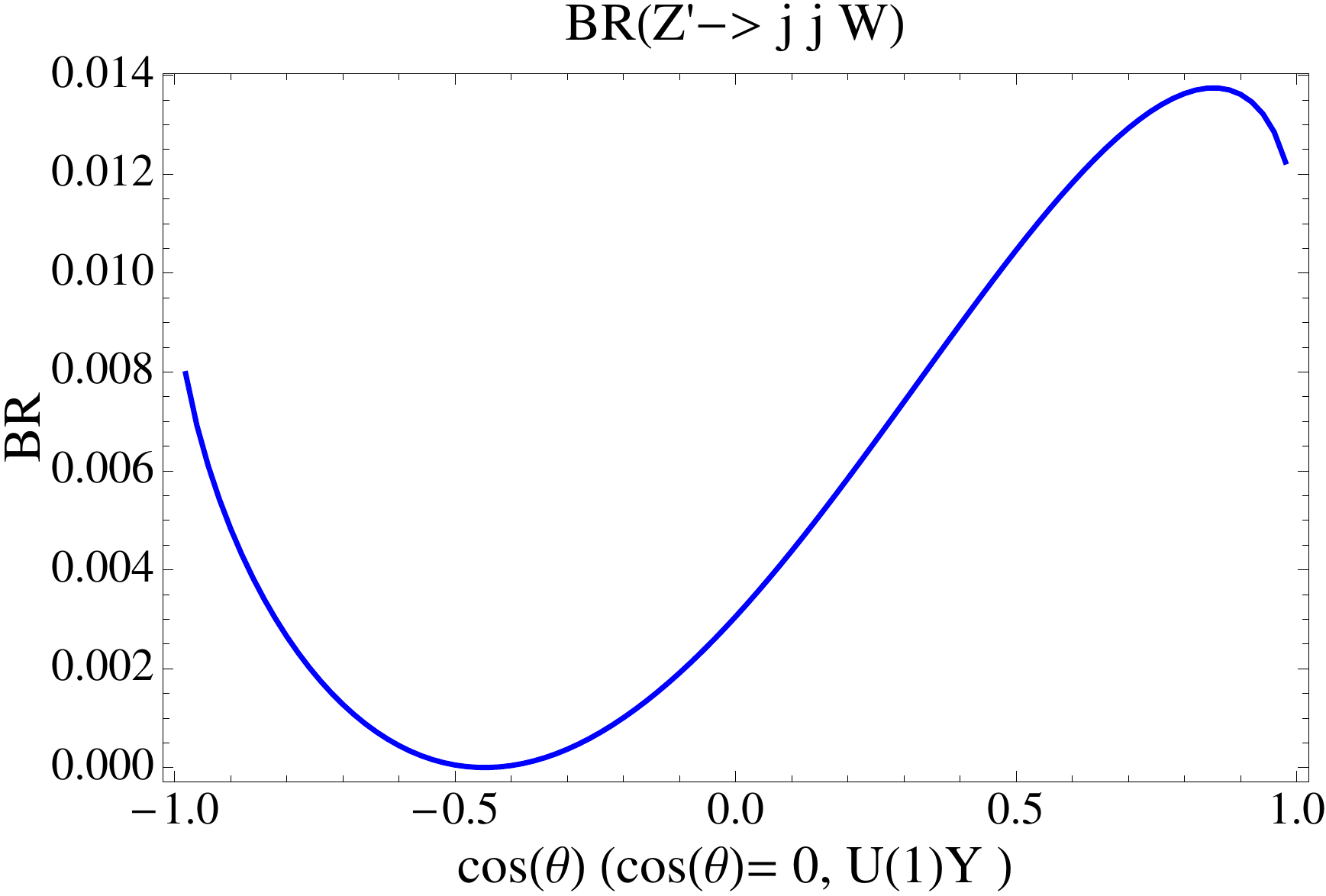}
\includegraphics[width=0.44\linewidth]{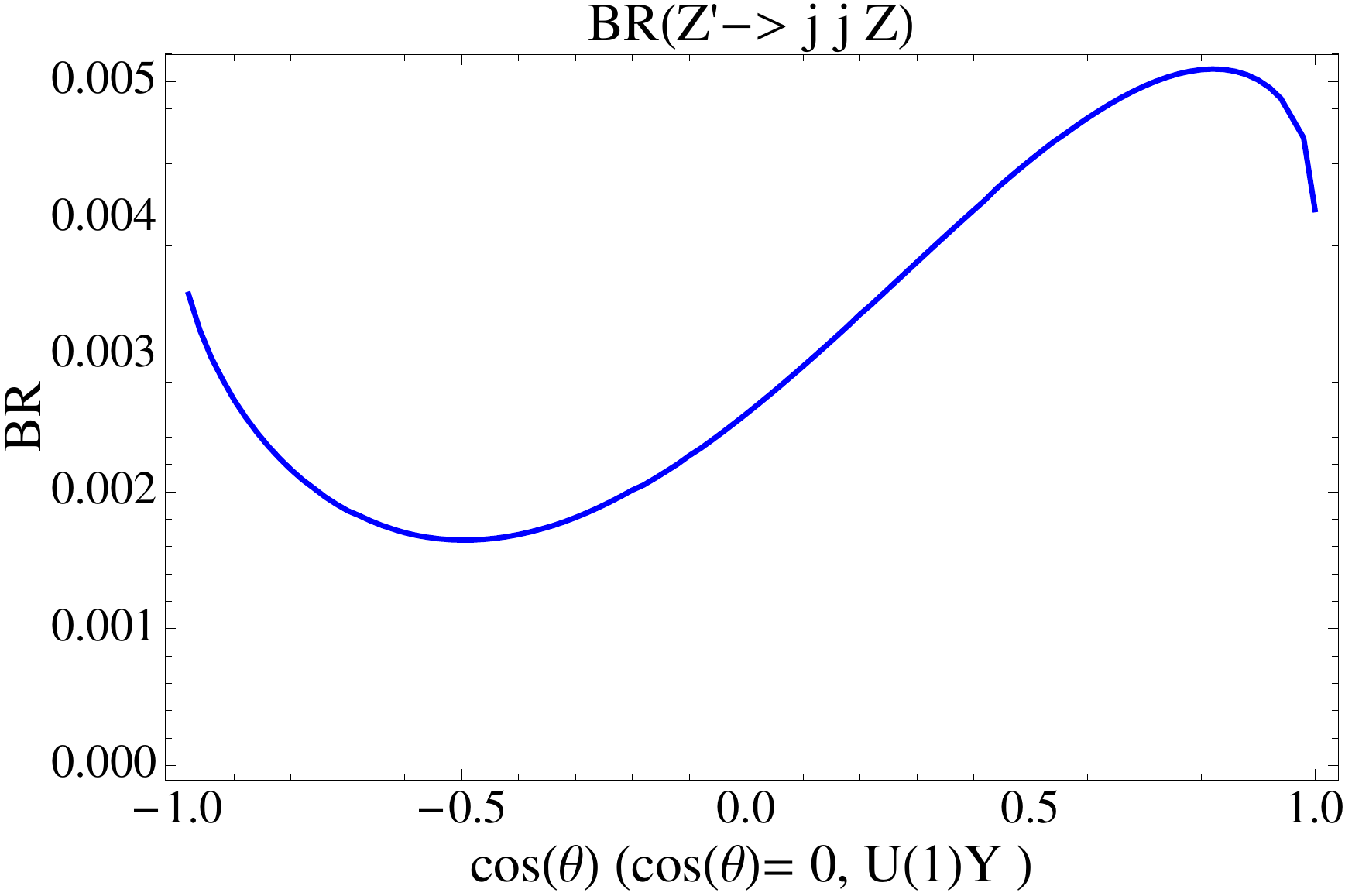}
\caption{ 3-body branching ratios for a 5 TeV $Z'$ particle, which include $W$ or $Z$ in final state.  We see that three body branching ratios can be large and that finding the invisible channel through radiation of a $Z$ or $W$ is possible.}  \label{Fig: Zprime2}
 \end{figure}

As shown in Fig.~\ref{Fig: Zprime2}, the invisible decay channel of the $Z'$ can be probed with 
EW radiation without suffering too large of a 
hit from branching ratios. 
The 3-body branching ratios are a few percent and can be seen.

 For leptonic final states, the ratio between the invisible decay channel $\nu \bar \nu Z$ and the 
leptonic decay channel $l^+ l^-$ allows one to measure $g_{l,L}^2 / (g_{l,R}^2 + g_{l,L}^2)$ where $g_{l,L}$ is the 
coupling of the $Z'$ to left handed leptons and $g_{l,R}$ is the coupling to right handed leptons.  
Thus the couplings of a $Z'$ to the SM chiral fermions can be measured at a 100~TeV hadron collider!

Although we do not study the details of $W$-tagging, 
we comment on the possibility of carrying out precision measurements using this potentially interesting tool. 
If $W$ tagging is efficient, it can also be used to probe the coupling of a $Z'$ to neutrinos.  The $W$ bosons couple to both left 
handed leptons and neutrinos with equal strength.  If the $Z'$ commutes with $SU(2)_L$, its couplings to the 
left handed leptons and neutrinos should be identical.   Its leptonic decays 
should radiate off $W$ bosons which are equally likely to appear 
collinear with the lepton or neutrino. On the other hand, a $Z'$ which does not  commute with $SU(2)_L$ would radiate $W$
bosons which preferentially align with either the lepton or neutrino.

$W$ tagging would also allow one to measure the decay channel $j j W$.  Comparison to the decay channel $j j$ 
gives a measurement of the ratio $g_{j,L}^2 / (g_{j,R}^2 + g_{j,L}^2)$.  While this measurement does not 
distinguish between the up and down type quarks, it does allow for a measurement of the chirality of the $Z'$ 
coupling to the quarks.  The decay channel $j j Z$ can also be used for a measurement, though the combination of both 
chiralities, up and down type quarks makes it less clean of a measurement.

\section{Quantum numbers from electroweak radiation}
\label{Sec:quantum}

As we have emphasized in Sec.~\ref{Sec:Zprime}, the radiation of electroweak gauge bosons can be used 
for precision measurements of new physics that might be discovered at the 100~TeV machine. That particular
example had to do with determining the chiral couplings of a new $Z'$.  In this section we further pursue this
approach showing how one can determine quantum numbers of new particles based on total EW gauge bosons emission.
Particles which are not charged under $SU(2)_L \times U(1)_Y$ do not radiate $W$ and $Z$
bosons and can thus be distinguished from their charged counterparts.  

We illustrate this effect in two examples, both assume discovery of SUSY at the TeV scale.    
In the first example we assume a ``natural SUSY'' - like spectrum at the TeV scale, namely 
a stop as an NLSP decaying into a neutralno LSP.
The left and right handed stops have different couplings to the $Z$.
 Due to electroweak symmetry breaking, they mix so that the NLSP is an admixture of two.  
We explore how electroweak Sudakov radiation of $Z$s can be used to bound the mixing angle of the NLSP stop. 

In the second example, we consider collider stable LSPs, 
which can also be a dark matter (DM) candidate.  
We show that when produced in SUSY cascades, 
Higgsino and wino LSPs have an order of magnitude more $W$ and $Z$ 
radiation as compared to the bino LSP. We also show that the higgsino-like neutralino (an $SU(2)_L$ doublet)
can be potentially differentiated from the wino-like neutralino (an $SU(2)_L$ triplet). 
 
\subsection{Stop mixing angle measurement}
\label{Sec:Stops}

SUSY with light third generation squarks is a well motivated~\cite{Dimopoulos:1995mi,Cohen:1996vb} 
and well studied scenario~\cite{Brust:2011tb,Essig:2011qg,Papucci:2011wy}. 
Such a spectrum has been traditionally motivated by the
naturalness problem.  
In the SM, only a few of the one-loop divergences to the Higgs mass are large.
Thus naturalness demands that only a handful of SUSY particles need to be light. These particles are the higgsinos, winos, gluinos, sbottoms,
and stops.  While all of the SUSY particles are important for solving the \emph{big hierarchy problem}, 
namely naturalness all the way to the Planck scale, they are not all important for canceling the largest one-loop 
divergences.  Therefore
a spectrum, where all of the other superpartners are heavier than the stops and sbottoms by factor of~10 or even more is still natural. 
   
Stops, which are responsible for canceling the top quadratic divergence to the Higgs mass, 
play a central and crucial role in this scenario.
While the LHC 
has a dedicated search program for 
this scenario~\cite{CMS:2013cfa,Chatrchyan:2013xna,Aad:2014qaa,ATLAS:2013cma}, the 
LHC14 reach for the stops decaying into a top and neutralino 
is limited to just little bit above 1~TeV.\footnote{In some other cases, e.g. RPV stops, this bound is much more 
modest~\cite{Bai:2013xla}.}
While a bound of $m_{\tilde t} > 1$~TeV would already suggest some tuning, it would be interesting to move 
forward and search for stops in the TeV range.  If we find stops at a TeV, we will be
discovering that the SM is fine tuned to the level of $\sim 1\%$, a level of coincidence that we have already 
seen in nuclear physics. 

The 100~TeV machine's reach will extend much further~\cite{Cohen:2013xda, Cohen:2014hxa}.
Here we show that for TeV scale stops the
100~TeV machine can give us valuable information about the chirality of the stops, bounding its mixing angle.

At large masses, the chirality of the stops can be measured by the additional radiation of a $Z$ or $W$ in the event.  
The  enhancement for the radiation of $Z$s and $W$s makes this measurement feasible at a 100~TeV 
machine.
Note however that the radiation of the EW gauge bosons from the stop is only \emph{single log enhanced} because the
collinear singularity in this case is cut off by the mass of the emitting particle (the stop) and effectively 
does not lead to any enhancement.   Meanwhile, both ISR and FSR have a Sudakov double log enhancement.  
Because both the decay products of the stop and the initial state quarks have the 
same chirality as the stop, the radiation strength provides a good measure of the chirality of the stop regardless of 
where the radiation came from.

To illustrate our point we choose three benchmark points: $m_{\tilde t} = 0.7$~TeV, $m_{\tilde t} = 1$~TeV and 
$m_{\tilde t} = 1.5$~TeV, all decaying into a massless neutralino.  Note that 
the first benchmark point can be easily discovered by the LHC
while the last one is inaccessible even for the LHC14.  
We do not consider heavier stops as their cross section is too small and the chirality measurement will probably not 
be feasible.

 \begin{figure}
  \centering
  \includegraphics[width=0.54\linewidth]{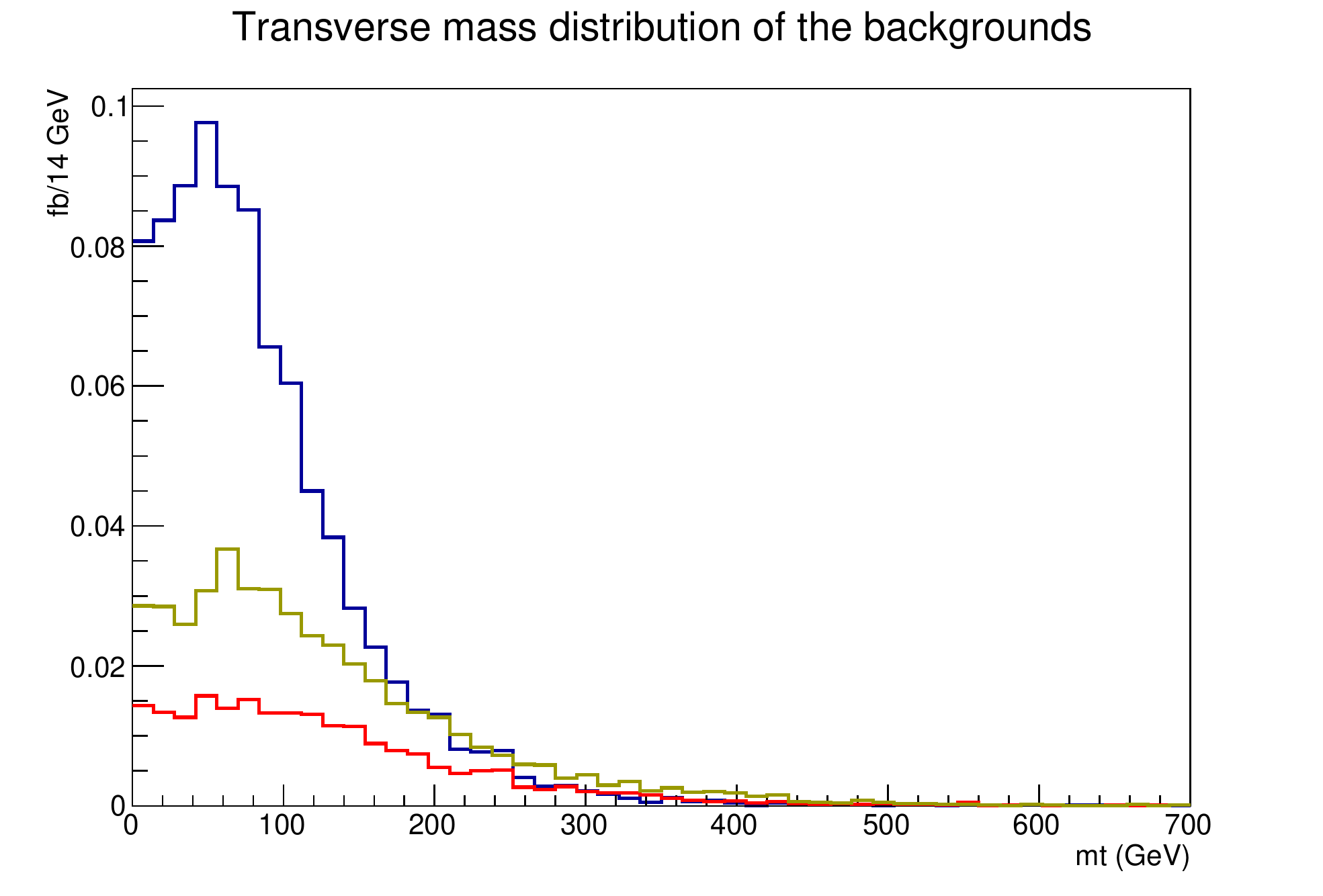}
  \caption{ $M_T$ distribution of the background calculated as explained in the text.
 The blue, red and yellow lines are the $t \bar t h$, $t \bar t Z Z$ and $ t \bar t W W$ background distributions 
respectively. We do not show $t \bar t Z$ since at the partonic level it cannot have $m_T$ in excess of $m_W$.}  
\label{Fig: MT}
 \end{figure}

Consider a stop NLSP decaying into a top quark and neutralino LSP. 
For simplicity we assume that the neutralino is bino-like,
however this technique can also be extended to other cases.
 To study the chirality of the NLSP stop, we propose
analyzing the production of $\tilde t \tilde t^* $ decaying into $t \bar t \chi^0 \chi^0$  accompanied by an emission of $Z$ at any 
stage of the process.   
We take advantage of an abundant semileptonic channel in $t \bar t$ final states and demand that the extra $Z$
decays leptonically. 

The list of possible 100~TeV backgrounds to this signal are: 
\begin{enumerate}
\item $t \bar t Z_l$ - by far the biggest cross section background but easily removed, 
since in the semileptonic channel naturally has $m_T(l, \met) < m_W$.
\item $t \bar t h$ with the Higgs decaying to $WW^*$, $ZZ^*$ or $\tau^+ \tau^-$.
\item $t \bar t Z_{ll} Z_\text{inv}$.  This background is smaller than the first background but is 
not immediately removed by a cut on the transverse mass of the lepton and $\met$.
\item $ t \bar t W_l W_l$ with leptons from $W$ reconstructing the $Z$ by chance.
\end{enumerate}

In order to remove these backgrounds, we first reconstruct the leptonic $Z$ in each of these events within 
a $m_Z \pm 15$~GeV window (if there is more than one candidate, we choose the candidate with the mass closest  
to $m_Z$). After that, we apply a transverse mass cut between the non-$Z$ lepton and $\met$.
As we can see from Fig.~\ref{Fig: MT}, a transverse mass cut can be very efficient in removing the background.  
We impose a transverse mass cut of 500~GeV to remove missing energy that stems from a $W$ while keeping the cuts on the leptons the same as in Sec.~\ref{Sec:ZWprime}.  After this cut, the main background is $ t \bar t W W$ 
with about 20 events at a luminosity of 3 ab$^{-1}$ (taking into account standard isolation and acceptance cuts).  

 \begin{figure}
 \centering
 \includegraphics[width=0.44\linewidth]{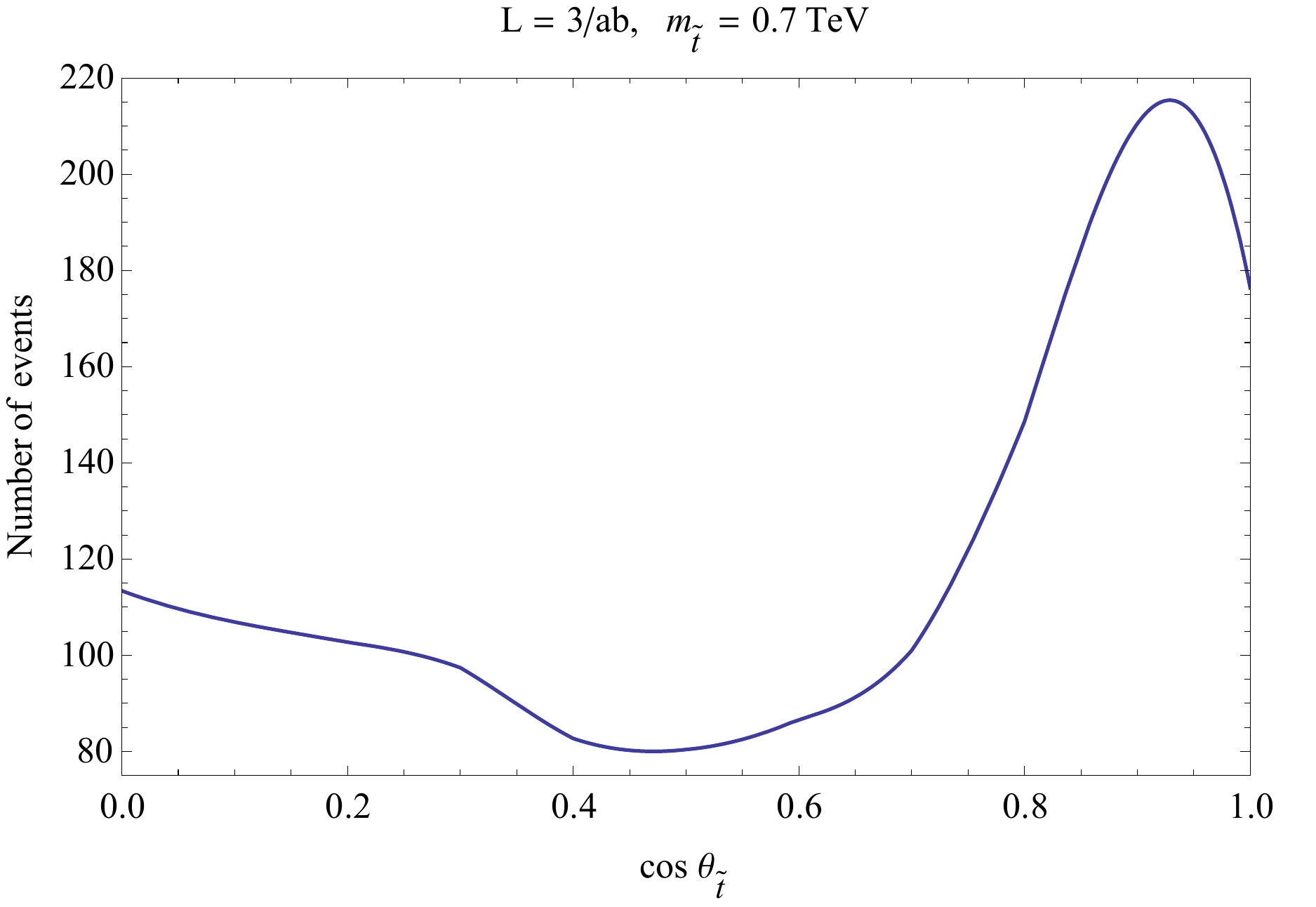}
 \includegraphics[width=0.44\linewidth]{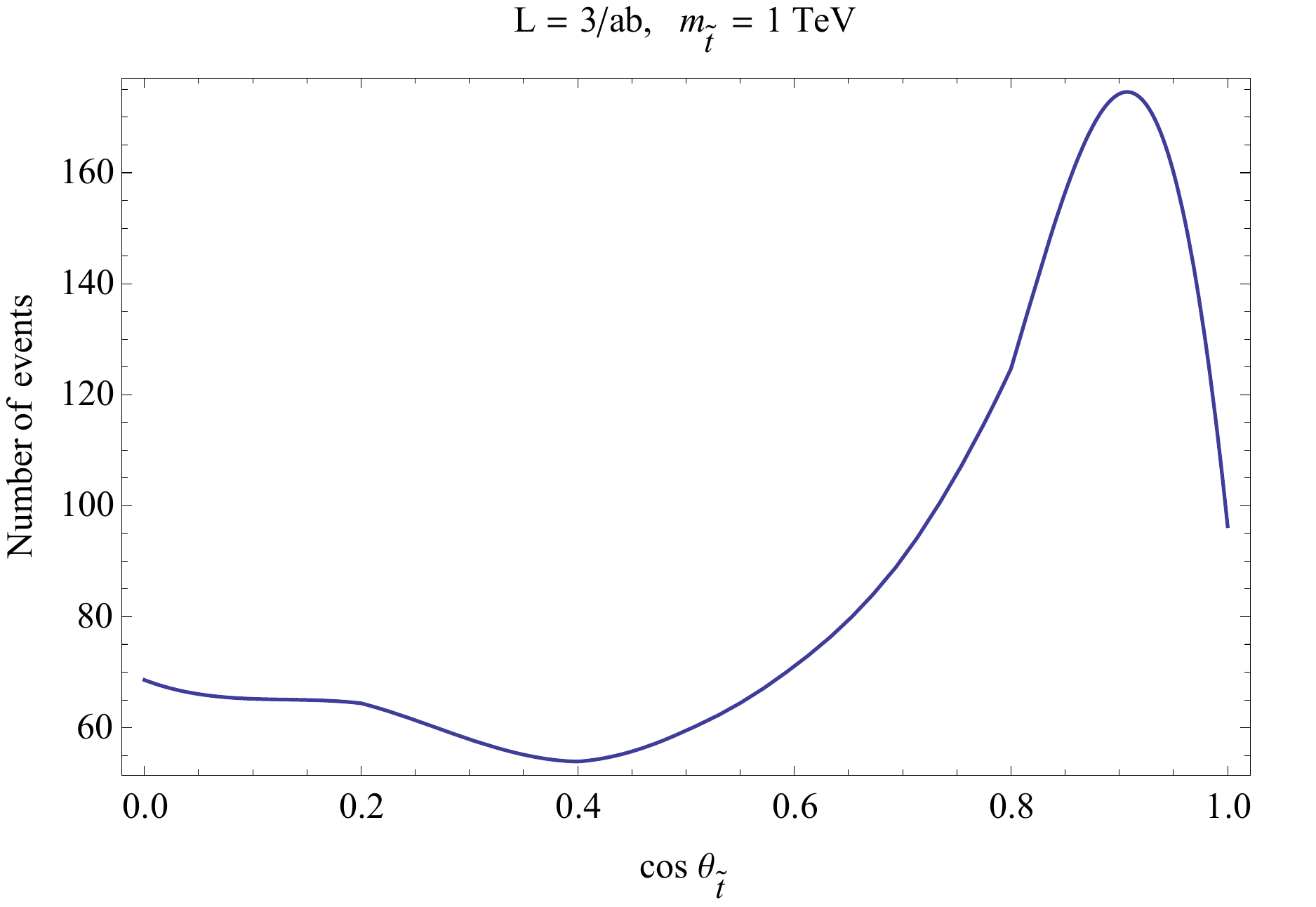}
  \includegraphics[width=0.44\linewidth]{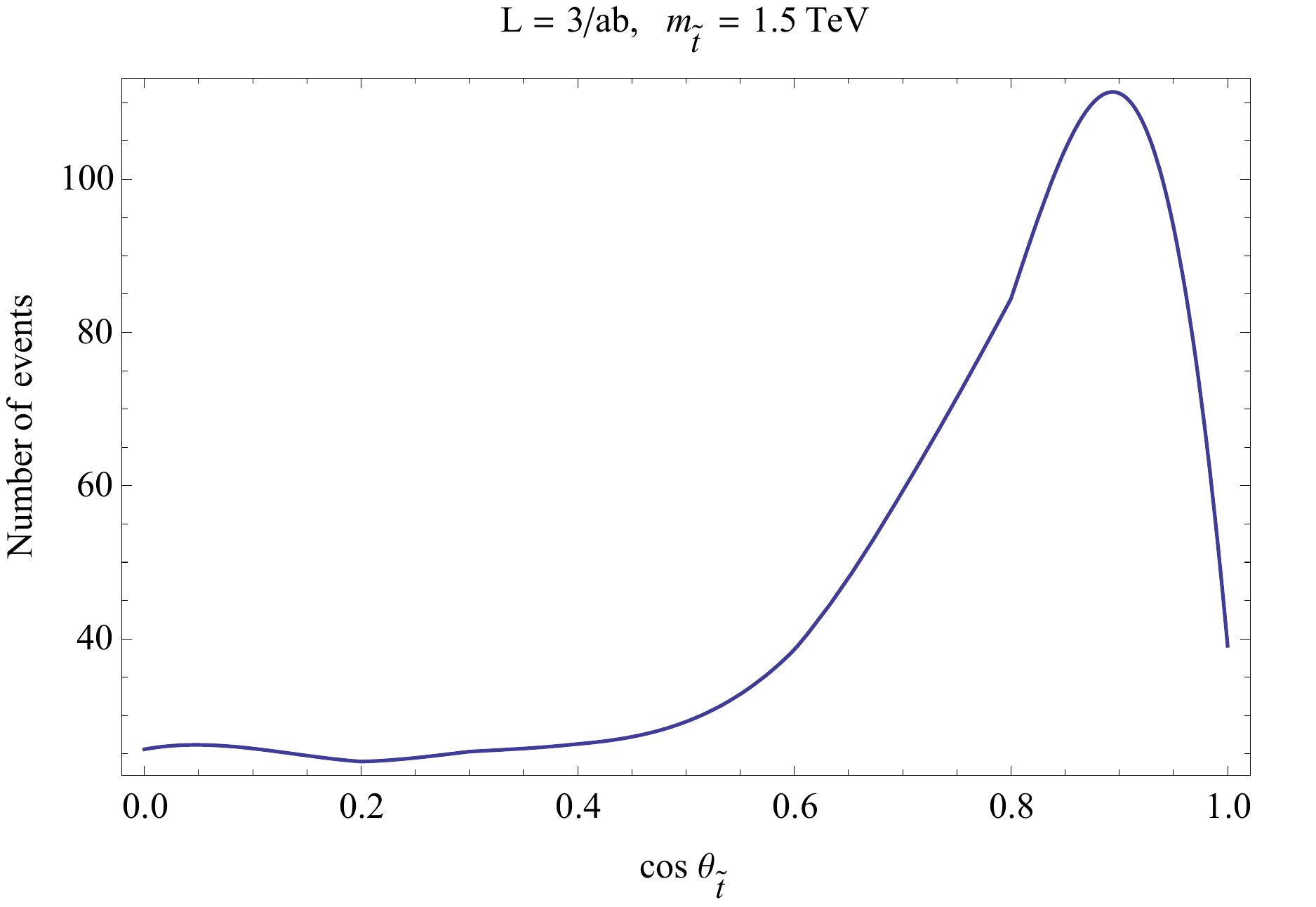}
 \caption{ Number of $\tilde t \tilde t^*$ events with an extra $Z_l$ passing the cuts as a function 
of $\cos\theta_{\tilde t}$ for three mass points.  $\cos\theta_{\tilde t} = 0$ is a right handed stop. Leptonic branching 
fractions of $Z$ have been taken into account. }  
\label{Fig: stops}
 \end{figure}

We show the predicted number of signal events as a function of mixing angle in Fig.~\ref{Fig: stops}.
We define the mixing angle between the stops as follow:
\beq
\left(
\begin{array}{c}
\tilde t_1\\
\tilde t_2
\end{array} \right) = 
\left( \begin{array}{cc}
\cos \theta_{\tilde t } & \sin \theta_{\tilde t} \\
- \sin \theta_{\tilde t} & \cos \theta_{\tilde t}
\end{array}
\right)
\left(
\begin{array}{c}
\tilde t_R \\
\tilde t_L
\end{array}
\right)~,
\eeq  
such that $\theta = 0$ corresponds to the lightest stop being purely right-handed.

There are several clear features observable from these plots.  The first is that there is a clear difference 
between $\cos\theta_{\tilde t} = 0$ and $1$.  Thus purely left and purely right handed stops can be 
distinguished.  A more surprising point is that the maximum/minimum of these plots are at values 
of $\cos\theta_{\tilde t}$ which is not 0 or 1.

  The maxima and minima are due to the emission of a $Z$ from the stop.  The emission of a 
	$Z$ from the top changes monotonically as a simple function of $\cos\theta_{\tilde t}$.  
	However, the emission of a $Z$ from the stop has a maximum away from $\cos\theta_{\tilde t} = 0$ or $1$.  
	The stop has $\theta_{\tilde t}$ dependent couplings to both the $Z$ and to the top and bino.  
	The coupling to the $Z$ is maximized in the purely left handed limit while the coupling to the bino is 
	maximized for the purely right handed limit.  The competition between these two terms 
	leads to the maximum at non-trivial $\cos\theta_{\tilde t}$.

Finally, the acceptances also change 
as a function of the mixing angle (due to differences in angular distributions of the leptons) and the mass 
of the stops. 
As the stop becomes heavier,
 more of the events pass the cuts.  
The change in acceptances also contributes to the variation in shape between the three masses choices.

\subsection{$SU(2)_L$ charge of the LSP}

In this subsection we show another example of a precision measurement one could perform at a 100~TeV collider 
if a more traditional SUSY spectrum is discovered at the TeV scale. Consider a spectrum with the gauginos 
lighter than the squarks. This spectrum can naturally arise since the the gaugino masses violate R-symmetry. If
the R-symmetry breaking is small compared to the SUSY breaking, one naturally gets this spectrum. In fact, very 
similar considerations led~\cite{ArkaniHamed:2004fb,Giudice:2004tc,ArkaniHamed:2004yi} 
to propose a split SUSY spectrum.  One can think about the spectrum that 
we analyze here as a minimal version of split SUSY (mini-split)~\cite{Arvanitaki:2012ps,ArkaniHamed:2012gw}.

In this case, SUSY production at the pp collider is dominated by $\tilde g \tilde g$, which further
cascade decays into the lightest neutralino through the off-shell third generation squarks (typically these are the lightest squarks
and dominate the cascade decays).   If this is the spectrum, it would be interesting to know what are the 
quantum numbers of the LSP. This question is especially interesting since the LSP is a DM candidate.  
The knowledge of whether the LSP is a singlet (bino), 
doublet (higgsino) or triplet (wino) of $SU(2)_L$ will also give us information about whether the LSP can be a thermal 
DM candidate. 

 \begin{figure}
  \centering
 \includegraphics[width=0.44\linewidth]{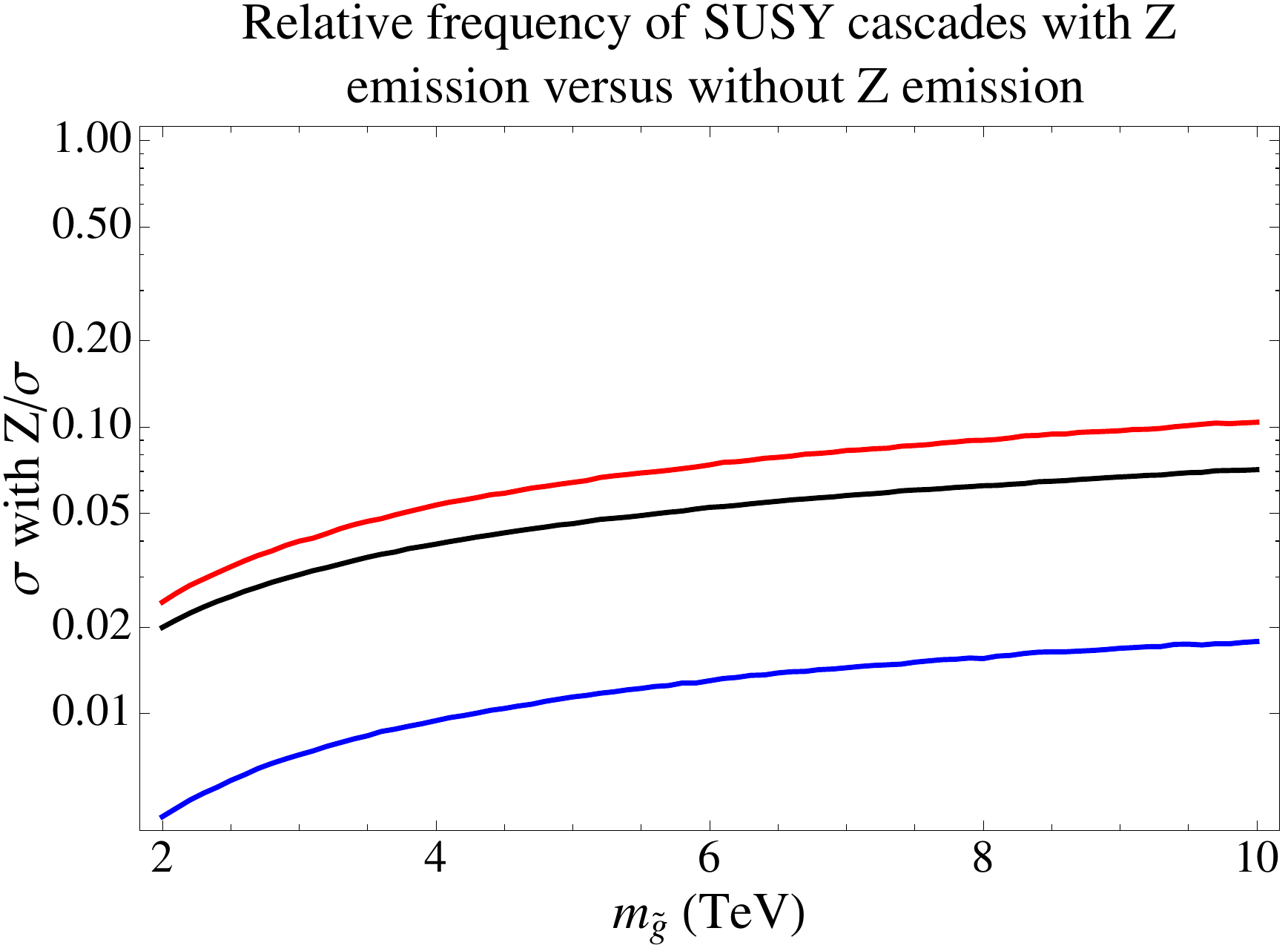}
 \includegraphics[width=0.44\linewidth]{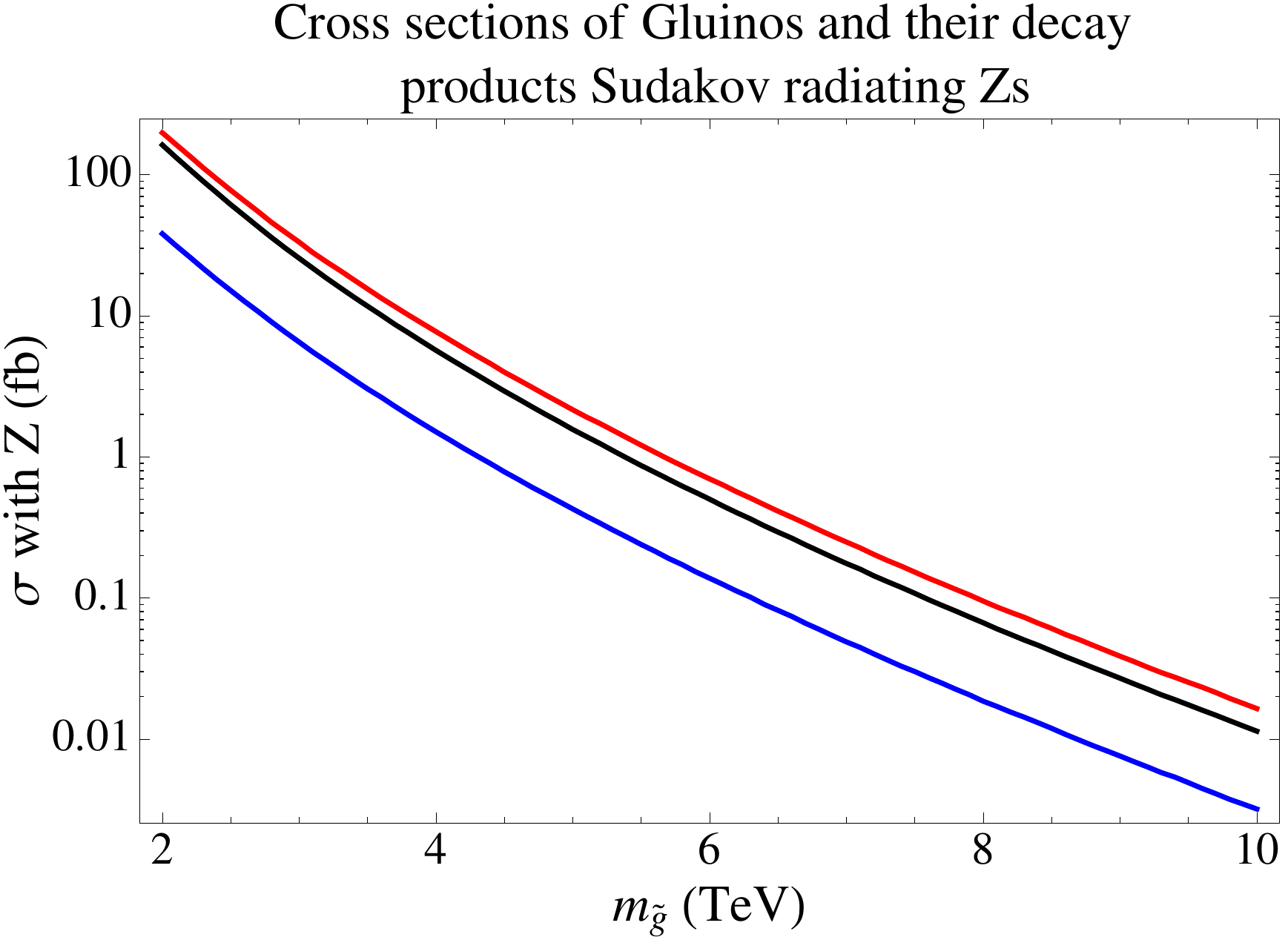}
  \caption{ Gluinos decay through stops and sbottoms 1.5 times its mass to varying LSPs.  The Sudakov enhancement is plotted as a function of Gluino mass.  The blue/black/red line indicate a bino/higgsino/wino LSP respectively.  Uncharged LSPs are very clearly distinguished from charged LSPs. Leptonic branching fractions of $Z$ have not been 
	taken into account. }  
	\label{Fig:gluino}
 \end{figure}

Here, we show that one can extract this information from the $Z$ radiation pattern in the gluino cascades. 
Of course, one could also ask whether this question can be answered by the production of $\chi^0 \chi^0+Z$. 
In fact the mono-$Z$ search for the DM has already been performed even at the 
LHC by ATLAS~\cite{Aad:2014vka,Aad:2013oja}
and it will
gain more ground in future. As we show in the Appendix, 
these searches will not discover the wino  
unless it is lighter than 1~TeV. If the LSP is an $SU(2)_L$ doublet, 
the result is expected to be even more modest. These mono-$Z$ searches are subdominant to the mono-jet 
searches~\cite{Low:2014cba} as the SUSY DM \emph{discovery channel}. 

In the split SUSY-like spectrum where the gluinos are pair produced and 
decay into the LSP through off-shell 3rd generation squarks, one can determine the 
$SU(2)_L$ quantum numbers of the LSP by simply considering the ratio
\beq
\frac{\sigma(\tilde g \tilde g + Z)}{\sigma(\tilde g \tilde g)}
\eeq 
where the $Z $ can be emitted \emph{at any stage of the production or decay}. 
We show that this ratio is sensitive to the quantum number of the LPS.  We consider the point in parameters space where
$m_{\tilde t} \approx m_{\tilde b} = 1.5 m_{\tilde g}$,  We calculate this ratio as well as the cross sections with the results shown
in Fig.~\ref{Fig:gluino}. For these plots we assume the LSP mass $m_{\tilde \chi^0} = 300$~GeV. The differences between the 
different LSP representations under $SU(2)_L$ are clearly visible. $Z$ emission from the singlet LSP is an order of 
 magnitude smaller than the other cases.  Higgsino emission is roughly a factor of 2 smaller 
than the wino emission. We are not trying to construct a search and estimate the possible reach, however 
the cross sections are large and the events are distinctive enough from the SM background events that 
we believe that a search along these lines can be implemented in a straight forward manner.

\section{Tests of unitarity}
\label{Sec:unitarity}

When the electroweak symmetry breaking sector is more complicated than the SM, the couplings of the Higgs boson to the rest of the SM particles changes.
These changes can lead to rather drastic consequences.  Consider a coupling of the Higgs to the EW gauge bosons 
of the form
\beq
\label{Eq: old}
\mathcal{L} \supset 2 a \frac{M_W^2}{v} h W^\mu W_\mu +  a \frac{M_Z^2}{v} h Z^\mu Z_\mu~
\eeq
$a=1$ in this formula corresponds to the well-known SM limit, where the $WW$ scattering is unitarized 
by the Higgs contribution. On the other hand, if $a \neq 1$, 
then the amplitude for longitudinal $WW$ scattering grows as 
\beq
{\cal A}(W_L W_L \rightarrow W_L W_L) \sim (1-a^2) E^2~.
\eeq
 At some point, $WW$ scattering becomes strongly coupled and necessitates the presence of new physics, which is 
responsible for taming this amplitude.  Some of the most well known examples of such models are Higgs 
compositeness~\cite{Georgi:1984af,Dugan:1984hq,Agashe:2004rs} and Higgs partial compositeness, 
see e.g.~\cite{Kaplan:1991dc,Gherghetta:2000qt}.  
From the point of view of a low energy effective theory, a complicated EWSB sector results in Higgs 
couplings which deviate from their SM values.  

In the past, several works have focused on the implications of deviations in the Higgs coupling to SM 
particles (see e.g.~\cite{Contino:2013gna,Contino:2011np}) using Eq.~\ref{Eq: old}.  The most common approach has 
been to study $WW$ scattering as its cross section can grow with energy squared.  
Traditional $WW$ scattering experiments should still be done at the 100~TeV machine and 
deserves a separate study.  

Rather than focusing on 
$WW$ scattering, 
in this section we focus on what happens when these deviations occur in the Higgs couplings to new physics.  
To motivate what type of new physics to consider, we consider an EWSB sector that couples to a new 
real scalar $S$:
\beq
\label{Eq:scalardecay}
\mathcal{L} = \lambda S (h^2 + a h_0^2 + 2 a h_+ h_- ) + \cdots
\eeq
At the first glance this coupling might look  unmotivated and contrived. However, scalars like this can easily arise 
in lots of new physics models.  The best example for such a scalar would be a two-Higgs doublet (2HDM)
in the decoupling 
and large $\tan \beta $ limit. In this case, the entire EWSB is in good approximation due to the low mass Higgs 
and the coupling of the heavy Higgs to the low energy EWSB sector 
is exactly of form~\eqref{Eq:scalardecay} with $a = 1$. 


If $a \ne 1$, then $WW$ scattering becomes strong at some energy scale due to the exchange of the scalar $S$, 
and is
\beq
{\cal A}( W^+W^- \to W^+ W^-) \propto \lambda^2 (1 -a^2) E^2~. 
\eeq  
As mentioned before, studying traditional $WW$ scattering at 100~TeV will 
place bounds on these deviations.

In addition to $WW$ scattering, the branching ratios for this new scalar $S$
 can give crucial information to the value of $a$.  If Eq.~\ref{Eq:scalardecay} is responsible for 
the decay of the scalar $S$, then unlike $WW$ scattering,  the branching ratios are \emph{independent} 
of the value of $\lambda$.  Thus if the couplings~\eqref{Eq:scalardecay} is present and 
$\lambda$ is small, branching ratios of $S$ are a better way to 
bound $a$ than $WW$ scattering is.

\begin{figure}
     \centering
     \includegraphics[width=0.5\linewidth]{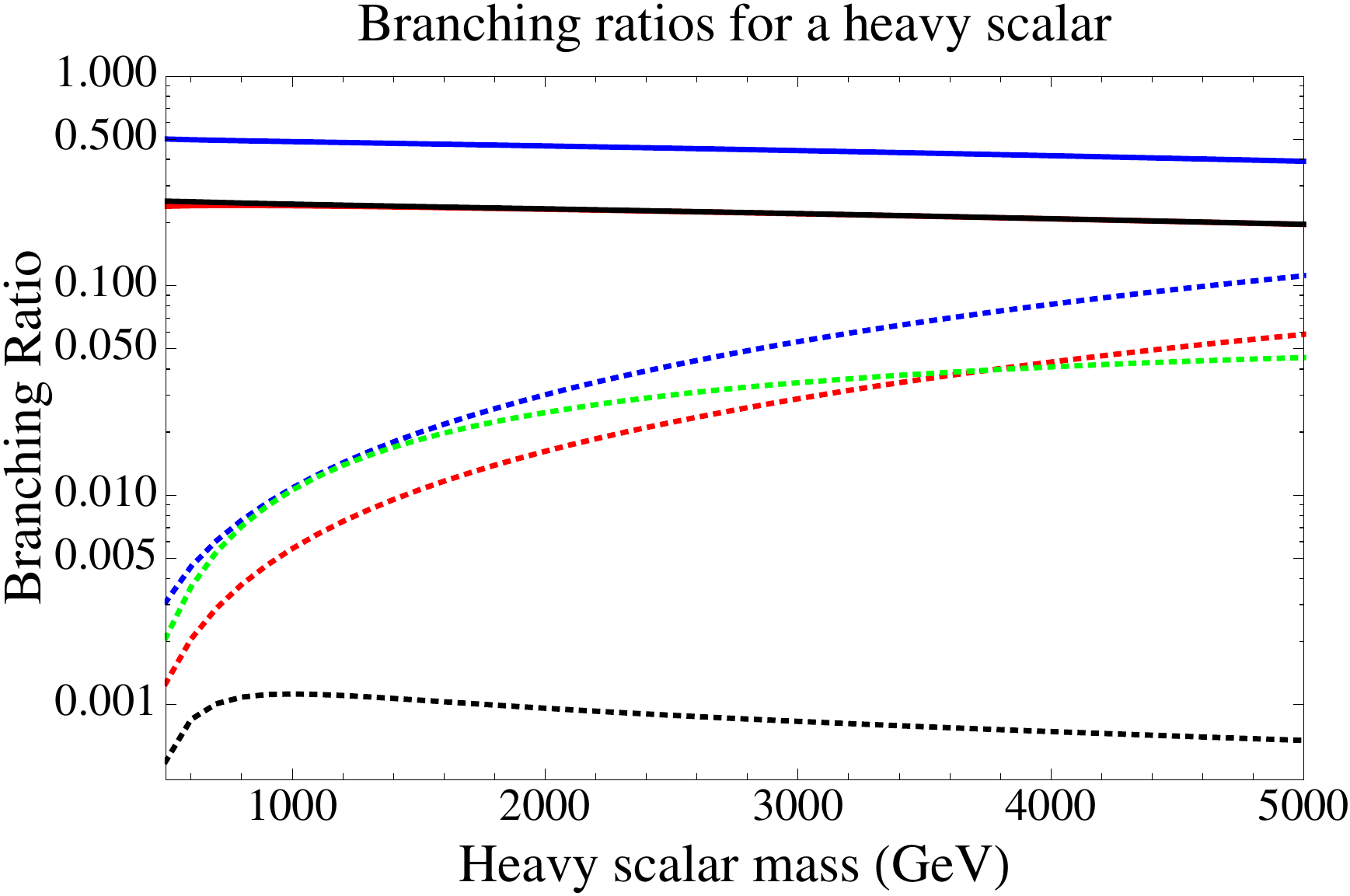}
     \caption{Branching ratios of a heavy scalar as a function of its mass.  The solid blue/red/black lines indicate the 2-body final states $WW/ ZZ/ hh$.  The dotted blue/ red/ green/ black lines indicate the 3-body final 
states $WWh/ ZZh/ WWZ/ hhh$.  
	The BRs are normalized to the sum of both two and three body decay widths.   }  \label{Fig: heavy decays}
 \end{figure}

We first consider the case when $a=1$.  Fig.~\ref{Fig: heavy decays} shows the branching 
ratios of $S$ as a function of its mass.  The two body branching ratios, $S \rightarrow WW/ZZ/hh$, 
are 2:1:1 respectively, as expected from the Goldstone boson equivalence theorem.  
In the three body decays, we see the standard Sudakov double logs.   
At large energies, the probability of a Higgs boson emitting a $W$ boson comes from the covariant derivative in the kinetic term
\beq
\label{Eq: higgs1}
\mathcal{L} \supset \partial^\mu h W^+_\mu h^-
\eeq
This coupling results in the standard EW Sudakov double log where a Higgs boson emits a transversely polarized 
$W^+$ and a Goldstone $h^-$.
The three body final states ($WWh$ and $ZZh$) have the standard 
double-log enhancements.  
 
We also see the difference between emitting a scalar versus a gauge boson.
Three scalar interactions, e.g. $hhh$, are relevant operators and are thus power law suppressed at large energies.  
Therefore we see that the $hhh$ branching ratio is significantly suppressed as compared to the 
other three body branching ratios.

\begin{figure}
\centering
\includegraphics[width=0.47\linewidth]{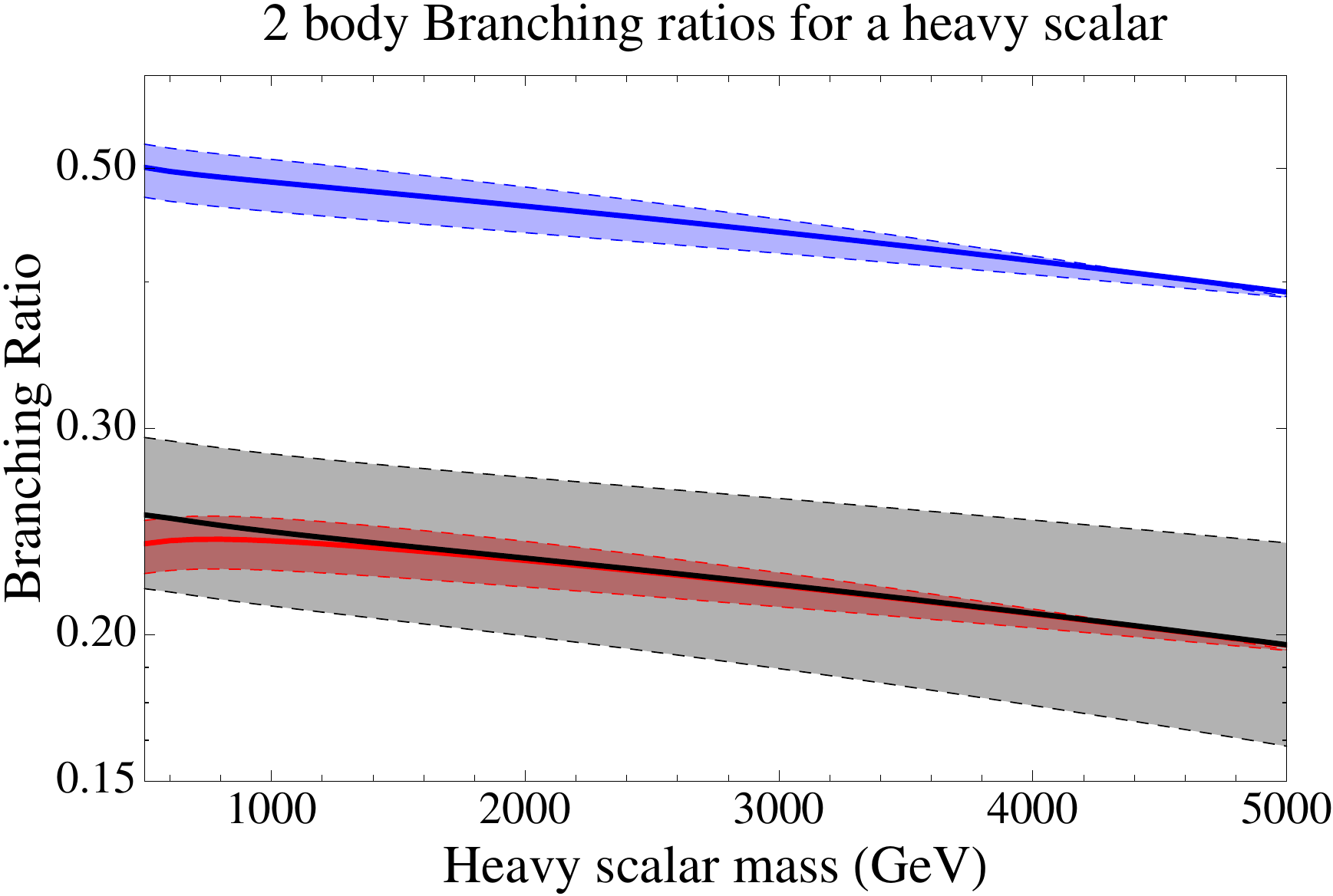}
\includegraphics[width=.47\linewidth]{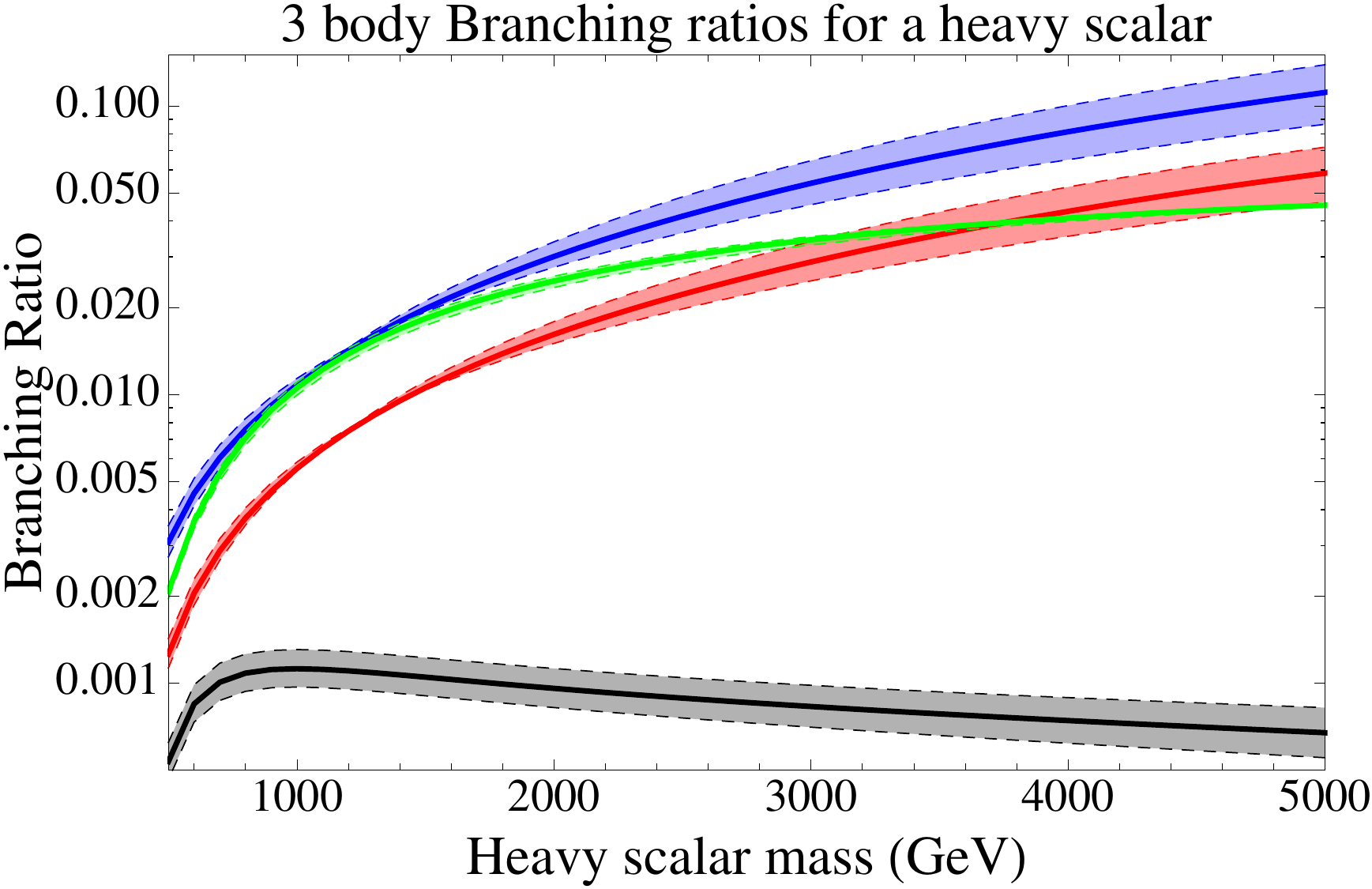}    
\caption{ Two and three body branching ratios for a scalar $S$ decaying through the term shown in 
Eq.~\eqref{Eq:scalardecay}.  On the left, the blue/red/black lines indicate the 2-body final states $WW/ZZ/hh$.  
On the right, the blue/red/green/black lines indicate the 3-body final states $WWh/ZZh/WWZ/hhh$.  
The shaded region indicates the region obtained by varying $a$ between $0.9$ and $1.1$.}  
\label{Fig: ane1}
 \end{figure}

If $a \ne 1$, both two-body and three-body decays are affected.  
The branching ratios in this case  are shown in Fig.~\ref{Fig: ane1}.  
Clearly, the first effect is that the 
two body branching ratios, $S \rightarrow WW/ZZ/hh$, no longer obey the 2:1:1 pattern 
as expected from the Goldstone boson equivalence theorem.  Except for very large masses, the two 
body branching ratios are larger than the three body branching ratios so that the deviation from 
the 2:1:1 pattern is the first effect to be observed.

However, the effect of $a \ne 1$ on three body decays is more dramatic.  If $a \ne 1$, then the term dominating 
EW radiation off of the Higgs is
\beq
\label{Eq: SM}
2 \frac{m_W^2}{v} h W^+ W^-
\eeq
Due to the bad behavior of the polarization vectors for the longitudinal component of the gauge boson, 
this term can potentially lead to poor behavior at high energies. If $a=1$, the poor high energy behavior 
of the two diagrams $S \rightarrow h h \rightarrow h WW$ and $S \rightarrow W W \rightarrow W W h$ 
cancel against each other.  But if $a\ne 1$, then the poor high energy behavior does not cancel, potentially allowing 
for much stronger growth of the three-body decay modes, which are tamed only by the new physics responsible 
for restoring unitarity.   

The shaded regions of Fig.~\ref{Fig: ane1} indicate the effect of varying $a$ between $0.9$ and $1.1$.  
We see that there are $\mathcal{O}(1)$ effects on the three body branching ratios.   
As the three body decays become more and more important, these effects become larger and larger.  Thus we see that by measuring three body branching ratios with $\mathcal{O}(1)$ uncertainties, we can bound $a$ by $\sim 10\%$.

\section{Conclusion}
\label{Sec:conclusion}

In this article we have sketched a few of conceptually  new searches which  will be  available at a 100~TeV pp collider.
These searches are based on a simple fact that the EW symmetry is effectively unbroken at the operation scale of 
a 100~TeV collider, meaning that these searches are largely impossible at the LHC. We considered three examples of
how different searches for new physics (SUSY, new gauge groups and compositeness) can be completely reshaped  
by the emission of EW $Z/W$ radiation. 

In this paper we have made many simplifying assumptions and all of
these searches, as well as many others, should be carefully reconsidered in future. 
First of all, in our simulations we worked only at the parton level and did not simulate showering
and hadronization effects. 
As the detectors for a 100~TeV collider are not know, we ignored all detector effects. 
Because of our simplifying assumptions, we ignored a lot of reducible backgrounds (e.g. $t\bar t Z$ 
in Sec.~\ref{Sec:Stops}), which might be extremely important and even dominant once all the effects 
of hadronization and detector simulation are properly included. We leave the crucial question of reducible 
backgrounds for future studies.  

We have also only included tree level effects.  There is a well known cancellation between real and virtual emission 
so that as the $p_T$ become larger these tree level calculations are 
not reliable~\cite{Moretti:2005ut,Moretti:2006ea,Dittmaier:2012kx,Stirling:2012ak}.  
Due these simplifying assumptions, our estimates are only accurate up to $\mathcal{O} (1)$ numbers.  
However, these effects are all grounded in physically significant effects and so we do not expect 
refinements in the estimates to overturn any of our results.\footnote{Note that in the QCD and QED 
one cannot rely on the leading order gluon/photon emission and in order to capture correctly the amount 
of emitted gauge bosons one should perform a resummation. Therefore, it would be a valid question to ask, 
whether one can rely on our results (even at the ${\cal O} (1) $ level) without a similar resummation in the EW 
sector. Of course, at arbitrarily high energies such resummation should be performed, however as we see from our 
results (see e.g. Fig.~\ref{Fig:gluino}), at a 100~TeV collider the effect is still of order 10\% or smaller, therefore 
the results are still reliable without a proper resummation. }

These results show that a 100 TeV collider is very different from a 14 TeV collider and 
that these differences go well beyond energy rescaling. 
Not only do calculations need to be done to more orders in perturbation theory, 
but qualitatively new phenomenon occur.  These new phenomenon allow one to further test the 
Standard Model and to devise new search strategies.  
As for new physics searches, these possibilities can potentially open a window into new intriguing 
\emph{precision measurements} at a 100~TeV collider. In this paper we showed how the invisible decay 
channel of a $Z'$ or the chirality of a stop can be estimated at 100~TeV, something which has been traditionally 
considered a motivation for linear colliders.

Finally we notice that the effect of enhanced $W/Z$ emission can potentially lead to lots of new and surprising 
searches that one cannot perform at smaller energies. These can be either searches for new physics or even the SM 
searches where we take advantage of the fact that we have a handle on particle polarization (not necessarily event 
by event, but at least statistically). Here are just several examples of searches along these lines, one can potentially consider:  
\begin{itemize}
\item Quark vs gluon jet tagging, using the fact that high-$p_T$ quarks can radiate EW bosons with an appreciable 
rate.   
\item Top polarization and polarized cross sections based on $W/Z$ emission.
\item $WW$ scattering both into 2 and 3-body final states. The later becomes progressively more important at 
high energies due to EW radiation. 
\item Measuring polarized parton distribution functions.
\end{itemize}

A 100 TeV collider is an exciting possibility with many new phenomena, what we have mentioned and considered 
here, is just the tip of the iceberg. 

 \acknowledgments 
We are especially grateful to Nima Arkani-Hamed for numerous suggestions and illuminating discussions, as well
as for collaboration in the early stages of this project.  
We are also grateful to Rafaele D'Agnolo, Patrick Draper, Graham Kribs and Matthew Reece 
for useful discussions.  We would like to thank Raffaele D'Agnolo
for helpful comments on the manuscript.  A.K. is grateful to the IAS for hospitality when part of the project has been done.  A.H. is supported by the DoE under contract DE-SC0009988.  A.K. is supported by the Center for the Fundamental Laws of Nature at Harvard.

\vspace{0.5cm}

{\bf Note added:} While this work was in preparation~\cite{Nima}, 
we became aware about the work of~\cite{Rizzo:2014xma}, 
which has a partial overlap with our results in Sec.~\ref{Sec:Zprime}. 

\appendix

\section{Determination of the DM quantum numbers with mono-$Z$ searches}

In Sec.~\ref{Sec:quantum}, we showed that in the context of a mini-split SUSY like spectrum, one can determine the $SU(2)_W \times U(1)_Y$
quantum numbers of the LSP by counting the events with an extra-$Z$ emitted in these events. In that particular 
case, we used the large gluino cross sections to boost the production and increase the number of events. However, 
one can also try to determine the quantum numbers of collider-stable particles 
when they are not accompanied by the boost in cross-section provided by the gluinos.  This question is especially relevant 
for DM, but is not limited to it. 

When DM is produced directly rather than in a cascade decay, we do not have anything to trigger on 
other than associated production with either jets or any other particle.  The discovery
channel even at 100~TeV collider will still be dominated by jet+$\met$ 
searches~\cite{Low:2014cba}.\footnote{For analogous LHC 
searches see~\cite{Aad:2014vka,Aad:2013oja}.}  
However, these searches are insensitive to the $SU(2)\times U(1)$ quantum 
numbers of the DM because the jet radiation is exclusively due to ISR.

\begin{figure}
\centering 
\includegraphics[width = .5\textwidth]{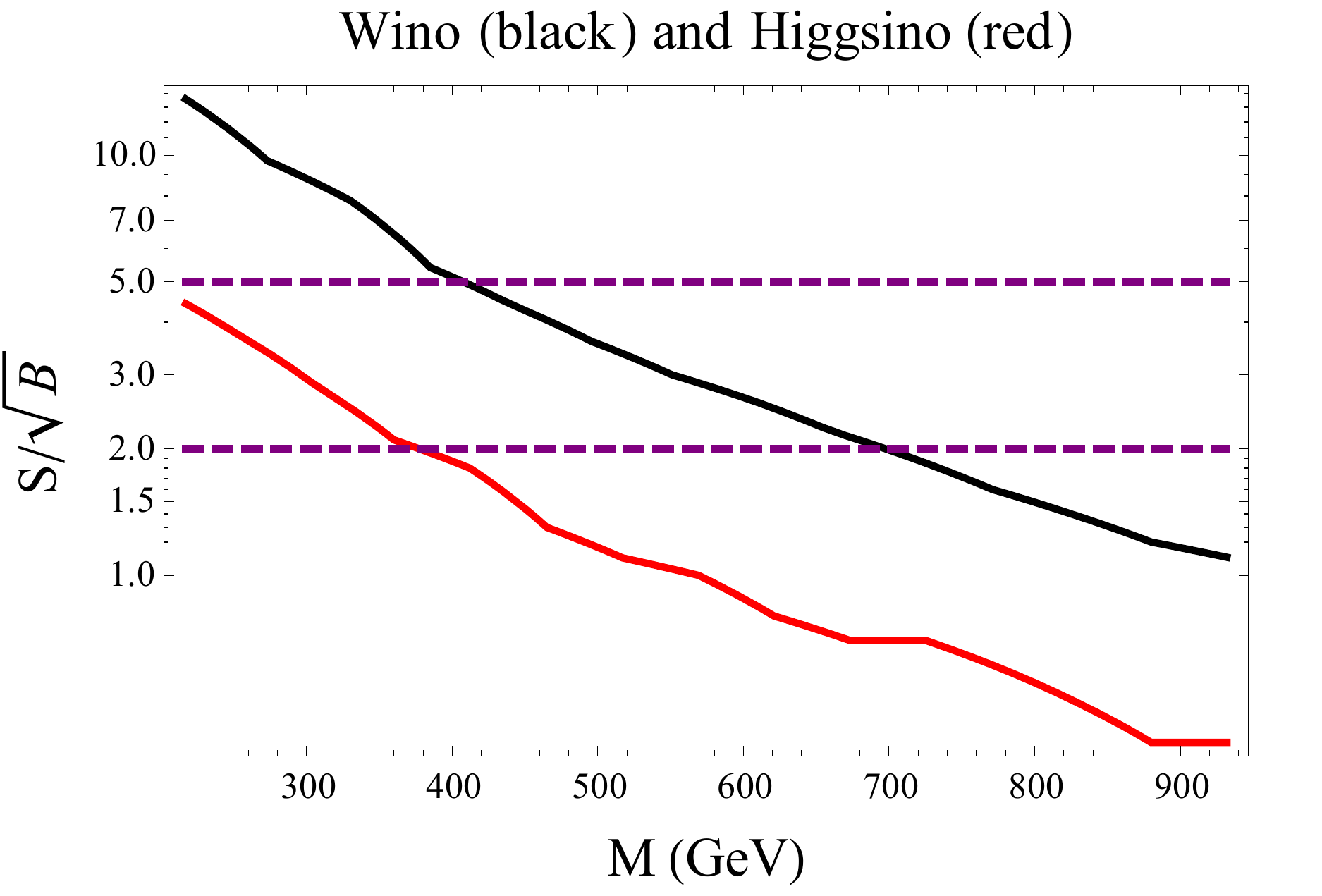}
\caption{Expected reach for SUSY dark matter with a $Z_l + \met$ search at 100~TeV collider with  
integrated luminosity ${\cal L} = 3$~ab$^{-1}$. }
\label{Fig:DMreach}
\end{figure}

In order to have any handle on DM's EW quantum numbers, 
we need to rely on the relatively rare channel of $Z_l + \met$.  The $Z$ can come from
FSR and is therefore expected to be different for $SU(2)$ triplets, doublets or singlets.  
The strength of the signal can be used to determine the EW quantum numbers of the DM.
As most DM particles will be produced near rest, the emission of Zs from DM will only be single log enhanced.

The dominant background is $Z_{inv} Z_l$.  We place a cut on a $p_T(Z_l)$ depending on the mass of the DM 
we are interested in. This cut varies from $p_T(Z_l) > 0.2$~TeV for the light DM 
(we start our scan from $m_{DM} = 200$~GeV)
and goes up to $p_T(Z_l) > 0.8$~TeV for the heaviest DM that we consider, which is 1~TeV.
The results of this search are shown in Fig.~\ref{Fig:DMreach}.  We see that because $Z$ 
emission is only single log enhanced,
the bounds are not competitive with the jet+$\met$ search (see Ref.~\cite{Low:2014cba}).
Even 
the winos (an $SU(2)_W$ triplet) cannot be excluded beyond 1~TeV using this channel, 
while the reach for higgsinos (an $SU(2)$
doublet) would reach 400~GeV at best.  We do not show the expected reach for the 
binos since the production rate for $Z_l + \met$
in this case is almost an order of magnitude smaller than for higgsinos, preventing any reasonable reach even at small
masses. 

We conclude that this search should be performed at 100~TeV collider, however it will be supplementary 
to the jet+$\met$ search and will give us useful information about the collider stable particle if its mass is below 1~TeV.

\bibliography{refs}
\bibliographystyle{jhep}

\end{document}